\documentclass[letterpaper,twocolumn,10pt]{article}
\usepackage[inchmargins, nocopyright]{sigmin}

\pdfpagewidth=\paperwidth
\pdfpageheight=\paperheight

\usepackage{times}
\usepackage{epsfig}
\usepackage{xspace}
\usepackage[dvipsnames]{xcolor}
\usepackage{hyperref}
\usepackage{listings}
\usepackage{wrapfig}
\usepackage[alwaysadjust]{paralist}
\usepackage{fancyhdr}

\usepackage{algorithm}
\usepackage[noend]{algpseudocode}

\makeatletter
\newcommand{\setalglineno}[1]{%
  \setcounter{ALG@line}{\numexpr#1-1}}
\makeatother

\algrenewcommand\algorithmicindent{0.75em}

\algblockdefx[AtomicBlock]{AtomicBlock}{EndAtomicBlock}[1][]{\textbf{atomic} #1}{\textbf{end atomic}}

\makeatletter
\renewcommand{\ALG@beginalgorithmic}{\small}
\makeatother

\newif\ifgray    % Flag for indicating whether code line should be grayed
\grayfalse
\renewcommand\alglinenumber[1]%
 {\ifgray\relax
  \noindent{\color{lightgray}\rule[-0.8ex]{\columnwidth}{3.1ex}\hbox{}\hskip-\columnwidth\hbox{}}%
  \fi
\footnotesize\arabic{ALG@line}}

\newlength{\continueindent}
\usepackage{etoolbox}
\makeatletter
\newcommand*{\ALG@customparshape}{\parshape 2 \leftmargin \linewidth \dimexpr\ALG@tlm+\continueindent\relax \dimexpr\linewidth+\leftmargin-\ALG@tlm-\continueindent\relax}
\apptocmd{\ALG@beginblock}{\ALG@customparshape}{}{\errmessage{failed to patch}}
\makeatother
\makeatletter
\providecommand\@enum@widestlabel{7}
\makeatother

\usepackage{xcolor}
\usepackage{graphicx}

\usepackage{amsmath}
\usepackage{amssymb}
\usepackage[caption=false,font=footnotesize,labelfont=sf,textfont=sf]{subfig}

\newcommand{\mypara}[1]{\vspace{4pt}\noindent\textbf{{#1}.}~}

\newcommand{\secref}[1]{Sec.~\ref{#1}\xspace}
\newcommand{\secsref}[2]{Secs.~\ref{#1}--\ref{#2}\xspace}
\newcommand{\figref}[1]{Fig.~\ref{#1}\xspace}
\newcommand{\figsref}[2]{Figs.~\ref{#1}--\ref{#2}\xspace}
\newcommand{\tblref}[1]{Table~\ref{#1}\xspace}
\newcommand{\figrefstatic}[1]{Fig.~{#1}\xspace}
\newcommand{\lineref}[1]{line~\ref{#1}\xspace}
\newcommand{\linesref}[2]{lines~\ref{#1}--\ref{#2}\xspace}
\newcommand{\appref}[1]{App.~\ref{#1}\xspace}

\newcommand{\gbytes}{\ensuremath{\mathrm{GB}}\xspace}
\newcommand{\mbytes}{\ensuremath{\mathrm{MB}}\xspace}

\newcommand{\ghertz}{\ensuremath{\mathrm{GHz}}\xspace}
\newcommand{\msecs}{\ensuremath{\mathrm{ms}}\xspace}

\newcommand{\secs}{\ensuremath{\mathrm{s}}\xspace}

\newcommand{\git}{\texttt{git}\xspace}
\newcommand{\gmail}{Gmail\xspace}
\newcommand{\xlib}{\texttt{Xlib}\xspace}
\newcommand{\xpilot}{{\it XPilot}\xspace}
\newcommand{\xpilotng}{{\it XPilot NG}\xspace}
\newcommand{\klee}{\textsc{klee}\xspace}
\newcommand{\cloudnine}{Cloud9\xspace}

\newcommand{\openssl}{OpenSSL\xspace}
\newcommand{\sslvE}{\texttt{1.0.1e}\xspace}
\newcommand{\sslvF}{\texttt{1.0.1f}\xspace}
\newcommand{\sslvEF}{\texttt{1.0.1ef}\xspace}
\newcommand{\sclient}{\texttt{s\_client}\xspace}
\newcommand{\sserver}{\texttt{s\_server}\xspace}
\newcommand{\heartbleed}{Heartbleed\xspace}

\newcommand{\tetrinet}{{\it TetriNet}\xspace}
\newcommand{\uclibc}{\texttt{uClibc}\xspace}
\newcommand{\ncurses}{\texttt{ncurses}\xspace}
\newcommand{\tetrinetTraces}{\ensuremath{20}\xspace}
\newcommand{\xpilotTraces}{\ensuremath{40}\xspace}

\newcommand{\msg}[1]{\ensuremath{\mathit{msg}_{#1}}\xspace}
\newcommand{\msgIdx}{\ensuremath{i}\xspace}

\newcommand{\msgNmbr}{\ensuremath{n}\xspace}

\newcommand{\posixSend}{{\tt send()}\xspace}
\newcommand{\posixRecv}{{\tt recv()}\xspace}

\newcommand{\sendInstr}{{\sc send}\xspace}
\newcommand{\recvInstr}{{\sc recv}\xspace}
\newcommand{\stdin}{\texttt{stdin}\xspace}
\newcommand{\execPrefix}[1]{\ensuremath{\Pi_{#1}}\xspace}
\newcommand{\execPrefixAlt}[1]{\ensuremath{\hat{\Pi}_{#1}}\xspace}
\newcommand{\symState}[1]{\ensuremath{\sigma_{#1}}\xspace}

\newcommand{\timestamp}{\ensuremath{t}\xspace}
\newcommand{\cost}[1]{\ensuremath{\mathit{cost}({#1})}\xspace}
\newcommand{\lag}[1]{\ensuremath{\mathit{lag}({#1})}\xspace}
\newcommand{\completion}[1]{\ensuremath{\mathit{comp}({#1})}\xspace}
\newcommand{\arrival}[1]{\ensuremath{\mathit{arr}({#1})}\xspace}

\newcommand{\Spawn}{\textbf{spawn}\xspace}

\newcommand{\Sync}{\textbf{sync}\xspace}

\newcommand{\node}{\ensuremath{\mathsf{nd}}\xspace}
\newcommand{\rootNode}{\ensuremath{\mathsf{Root}}\xspace}
\newcommand{\resultNode}{\ensuremath{\mathsf{Rslt}}\xspace}
\newcommand{\Node}{\ensuremath{\mathsf{Node}}\xspace}

\newcommand{\childField}[1]{\ensuremath{\mathsf{child}_{#1}}\xspace}
\newcommand{\pathField}{\ensuremath{\mathsf{path}}\xspace}
\newcommand{\stateField}{\ensuremath{\mathsf{state}}\xspace}
\newcommand{\prevConstraintsField}{\ensuremath{\mathsf{saved}}\xspace}
\newcommand{\constraints}{\ensuremath{\mathsf{cons}}\xspace}
\newcommand{\isSymbolicBranch}{\ensuremath{\mathsf{isSymbolicBranch}}\xspace}
\newcommand{\isIOInstruction}{\ensuremath{\mathsf{isNetInstr}}\xspace}
\newcommand{\isProhibitive}{\ensuremath{\mathsf{isProhibitive}}\xspace}
\newcommand{\isNormal}{\ensuremath{\mathsf{isNormal}}\xspace}
\newcommand{\newState}{\ensuremath{\sigma}\xspace}
\newcommand{\newStateAlt}{\ensuremath{\sigma'}\xspace}
\newcommand{\newPath}{\ensuremath{\pi}\xspace}
\newcommand{\nextInstruction}{\ensuremath{\mathsf{nxt}}\xspace}
\newcommand{\execStep}{\ensuremath{\mathsf{execStep}}\xspace}
\newcommand{\execSymbolicSkip}{\ensuremath{\mathsf{execStepProhibitive}}\xspace}
\newcommand{\makeNode}{\ensuremath{\mathsf{makeNode}}\xspace}
\newcommand{\clone}{\ensuremath{\mathsf{clone}}\xspace}
\newcommand{\messageVar}{\ensuremath{\mathsf{msg}}\xspace}
\newcommand{\condition}{\ensuremath{\mathsf{cond}}\xspace}

\newcommand{\readyQ}{\ensuremath{\mathsf{Q}_{R}}\xspace}
\newcommand{\addedQ}{\ensuremath{\mathsf{Q}_{A}}\xspace}
\newcommand{\verifyFinished}{\ensuremath{\mathsf{Done}}\xspace}

\newcommand{\workerCount}{\ensuremath{\mathsf{NumWorkers}}\xspace}
\newcommand{\makeNodeQueue}{\ensuremath{\mathsf{makeNodeQueue}}\xspace}

\newcommand{\enqueue}{\ensuremath{\mathsf{enqueue}}\xspace}

\newcommand{\tryDequeue}{\ensuremath{\mathsf{dequeue}}\xspace}
\newcommand{\parallelVerifyAlg}{\ensuremath{\mathsf{ParallelVerify}}\xspace}
\newcommand{\nodeScheduler}{\ensuremath{\mathsf{NodeScheduler}}\xspace}
\newcommand{\selectNode}{\ensuremath{\mathsf{SelectNode}}\xspace}
\newcommand{\verifyWorker}{\ensuremath{\mathsf{VfyMsg}}\xspace}

\newcommand{\codeFalse}{\ensuremath{\textbf{false}}\xspace}
\newcommand{\codeTrue}{\ensuremath{\textbf{true}}\xspace}

\begin{document}
\date{}
\title{\Large \bf Server-side Verification of Client Behavior in Cryptographic Protocols}

\author{\begin{tabular}{ccccc}
    Andrew Chi & Robert Cochran & Marie Nesfield & Michael K.\ Reiter & Cynthia Sturton\\[10pt]
    \multicolumn{5}{c}{University of North Carolina} \\
    \multicolumn{5}{c}{Chapel Hill, NC, USA}
    \end{tabular}}

\maketitle
\thispagestyle{empty}

\subsection*{Abstract}
Numerous exploits of client-server protocols and applications involve
modifying clients to behave in ways that untampered clients would not,
such as crafting malicious packets.  In this paper, we demonstrate
practical verification of a cryptographic protocol client's messaging
behavior as being consistent with the client program it is believed to
be running. Moreover, we accomplish this without modifying the client
in any way, and without knowing all of the client-side inputs driving
its behavior. Our toolchain for verifying a client's messages explores
multiple candidate execution paths in the client concurrently, an
innovation that we show is both specifically useful for cryptographic
protocol clients and more generally useful for client applications of
other types, as well.  In addition, our toolchain includes a novel
approach to symbolically executing the client software in multiple
passes that defers expensive functions until their inputs can be
inferred and concretized.  We demonstrate client verification on
\openssl to show that, e.g., \heartbleed exploits can be detected
without \heartbleed-specific filtering and within seconds of the first
malicious packet, and that verification of legitimate clients can keep
pace with, e.g., \gmail workloads.

\section{Introduction}
\label{sec:intro}

Tampering with clients in client-server protocols or applications is
an ingredient in numerous abuses.  These abuses can involve exploits
on the server directly, or manipulation of application state for which
the client is authoritative.  Examples of the former include at least
ten vulnerabilities in the last two years for \openssl
alone\footnote{CVE-2014-0160, CVE-2014-3512, CVE-2014-3567,
  CVE-2014-3513, CVE-2015-0205, CVE-2015-1787, CVE-2015-0293,
  CVE-2015-0292, CVE-2015-0208, CVE-2015-0291.}, including the
high-profile Heartbleed~\cite{durumeric14:heartbleed} vulnerability,
which enabled a tampered SSL client to extract contents of server
memory.  Examples of the latter are ``invalid command'' game cheats
that permit the client greater powers or resources in the
game~\cite{webb08:survey}.

The ideal defense would be to implement formally verified servers that
incorporate all necessary input validation and application-specific
checking.  However, in practice, current production servers have
codebases too large to retrofit into a formally verified model.
Moreover, even simple input validation is difficult to get perfect,
despite extensive review, as exhibited by widely deployed security
software such as implementations of Transport Layer Security (TLS).
In 2014, critical vulnerabilities were discovered in all five major
implementations of TLS~\cite{leyden14:tlspwned}, many of which turned
out (after the fact) to be failures of input validation.

\begin{figure}[hbt]
\centering
\includegraphics[width=\columnwidth]{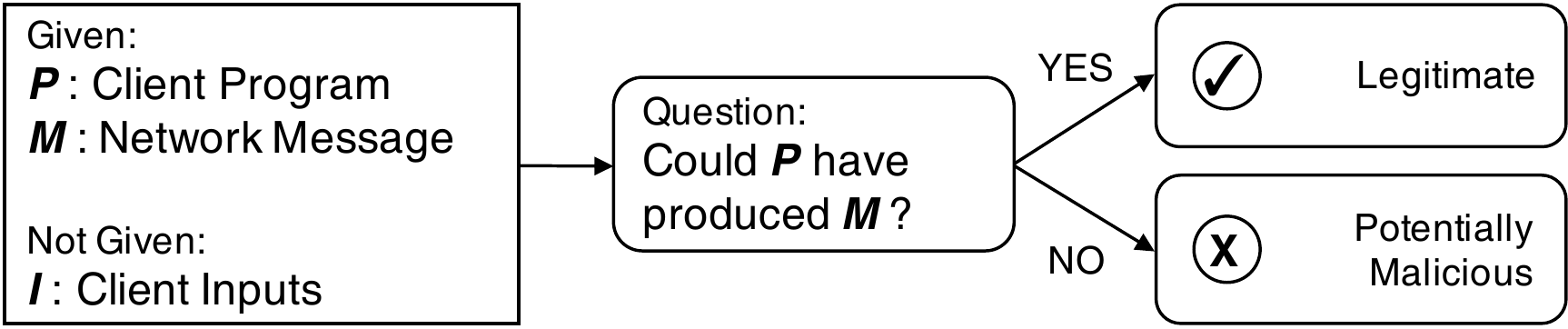}\\
\caption{Abstracted client verification problem.}\label{fig:aprob}
\end{figure}

Since it is generally impossible to anticipate all such abuses, in
this paper we explore a holistic approach to validating client
behavior as consistent with a sanctioned client program.  In this
approach, a \textit{verifier} monitors each client message as it is
delivered to the server, to determine whether the sequence of messages
received from the client so far is consistent with the program the
client is believed to be running and the messages that the server has
sent to the client (\figref{fig:aprob}).  Performing this verification
is challenging primarily because inputs or nondeterministic events at
the client may be unknown to the verifier, and thus, the
verifier must solve for whether there exist inputs that could have
driven the client software to send the messages it did.  Furthermore,
some of those inputs may be protected by cryptographic guarantees
(private keys in asymmetric cryptography), and maliciously crafted
fields may themselves be hidden by encryption, as is the
case with \heartbleed.

Our central contribution is to show that
legitimate cryptographic client behavior can in fact
be verified, not against a simplified protocol model but against
the client program.  Intuitively, we limit an attacker to only
behaviors that could be effected by a legitimate client.
We believe this advance to be important: in
showing that messages from a client can be quickly verified as
legitimate or potentially malicious, we narrow the time
between zero-day exploit and detection/countermeasure to minutes or
even seconds.  This is significant, since in the case of \heartbleed,
for example, the bug was introduced in March 2012 and disclosed in
April 2014, a window of vulnerability of over two years.  In fact,
with latency-tolerant applications (e.g., SMTP), our defense could be
installed inline as a \textit{prevention} system rather than simply a
detection system.  Moreover, our technique accomplishes
this verification with no vulnerability-specific configuration and,
indeed, would discover client exploit attempts even prior to a
vulnerability's disclosure.  It also requires no changes to client or
server software.\footnote{In the case of validating \openssl client
  behavior, we do use a common diagnostic feature on servers: logging
  session keys to enable analysis of network captures.}

Following several other works in verification of client messages when
some client-side values are unknown (see \secref{sec:related} for a
discussion of prior research), the basic strategy we take is to use
symbolic execution~\cite{boyer75:select} to trace the client execution
based on the messages received so far from the client (and the
messages the server has sent to it).  When the verifier, in tracing
client execution, operates on a value that it does not know (a
``symbolic'' value), it considers all possibilities for that value
(e.g., branching in both directions if a branch statement involves a
symbolic variable) and records constraints on those symbolic values
implied by the execution path taken.  Upon an execution path reaching
a message send point in the client software, the verifier reconciles
the accumulated constraints on that execution path with the next
message received from the client. If the path does not contradict
the message, then the message is confirmed as being consistent with
some valid client execution.

Our technical innovations within this direction of research are
twofold.
\begin{enumerate}
\iftrue
\item Prior research on this form of client verification
  has primarily focused on carefully prioritizing candidate
  paths through the client in the hopes of finding one quickly to
  validate the message trace observed so far.  This prioritization can
  itself be somewhat expensive (e.g., involving edit-distance
  computations on execution paths) and prone to error, in which case
  the verifier's search costs grow dramatically
  (e.g.,~\cite{cochran13:verification}).  Here we instead use
  parallelism to explore candidate paths concurrently, in lieu of
  sophisticated path prediction or to overcome the impact of poor
  predictions. 
\else
\item The performance of client verification hinges critically on
  carefully prioritizing the exploration of execution paths in the
  client program toward paths likely to confirm the validity of the
  message trace observed so far.  Despite prior research on quickly
  inferring candidate paths (e.g.,~\cite{cochran13:verification}), the
  primary obstacle to keeping pace with highly interactive
  applications remains the prediction of execution fragments to guide
  the verifier's search that end up being poor choices, in which case
  the search for an execution path to ``explain'' this message can
  grow substantially.  To dampen the impact of poor predictions, our
  verifier explores the possible paths through the client software in
  parallel.  
\fi
  In \secref{sec:parallel}, we describe the architecture of
  this parallel client-verification tool and demonstrate its ability
  to overcome limitations in path prediction accuracy on two online
  games studied in previous research.  We also highlight one aspect of
  cryptographic protocols for which efficiently validating client-side
  behavior depends on being able to explore multiple execution
  fragments in parallel, namely execution fragments reflecting
  plaintexts of different sizes, when the true plaintext size is
  hidden by message padding (as in SSH and draft TLS 1.3).  In this case,
  predicting the plaintext length is not possible from the ciphertext
  length, by design, and so exploring different candidate lengths in
  parallel yields substantial savings.

\item When verifying the behavior of a client in a cryptographic
  protocol such as Transport Layer Security (TLS), the search for a
  client execution path to explain the next client message can be
  stymied by paths that contain cryptographic functions for which some
  inputs are unknown (i.e., symbolic).  The symbolic execution of,
  e.g., the AES block cipher with an unknown message or a modular
  exponentiation with an unknown exponent is simply too costly.  Every
  message-dependent branch in the AES code or modular exponentiation
  routine would need to be explored, and the resulting formula would
  be unmanageably complex.  In \secref{sec:multipass} we therefore
  describe a multi-pass algorithm for exploring such paths, whereby
  user-specified ``prohibitive'' functions are bypassed temporarily
  until their inputs can be deduced through reconciliation with the
  client message; only then is the function explored (concretely).  In
  cases where those inputs can never be inferred---as would be the
  case for an ephemeral Diffie-Hellman key, for example---the system
  outputs the assumption required for the verification of the client
  message to be correct, which can be discharged from a small
  whitelist of assumptions.  Aside from these assumptions, our
  verification is sound.
\end{enumerate}
  
We stress that our verification technique, while not completely
turnkey, does not require detailed knowledge of the protocol or
application being verified.  For example, the specification of
prohibitive functions and a matching whitelist of permissible
assumptions is straightforward in our examples: the forbidden
functions are simply the AES block cipher, hash functions, and
elliptic curve group operations; and the whitelist includes the
assumptions that a group element can be produced by raising the group
generator to some power and that there exists an input that would
induce the hash function to produce a given value (which are both
reasonable).  Aside from specifying the forbidden functions and the
whitelist, the other (optional) step is to ``stub out'' calls to
software that are irrelevant to the analysis (e.g., \texttt{printf}).
For each of the contributions above, we detail the effort required by
the user to utilize them.

We document the efficacy of our client verification technique by
showing that it can keep pace with client messages in an interactive
\gmail session running over TLS 1.2 connections.  Our verification for
each client-to-server TLS record takes an average of 126\msecs on a
3.2\ghertz processor.  Taking into account the bursts of network
activity in \gmail traffic, and given that a message cannot begin
verification until all previous messages are verified, the
verification of any client-to-server message completes a maximum of
14\secs after the time the packet was delivered over the network.  We
also show that our technique keeps pace with TLS connections that use
message padding, a draft TLS 1.3~\cite{tls13} feature that introduces
costs that our parallel approach is able to overcome.

\section{Related Work}
\label{sec:related}

The most closely related work is due to Bethea et
al.~\cite{bethea11:games} and Cochran and
Reiter~\cite{cochran13:verification}.  These works develop algorithms
to verify the behavior of (non-cryptographic) client applications in
client-server settings, as we do here.  Bethea et al.\ adopted a
wholly offline strategy, owing to the expense of their techniques.
Cochran and Reiter improved the method by which a verifier searches
for a path through the client program that is consistent with the
messages seen by the verifier so far.  By leveraging a training phase
and using observed messages to provide hints as to the client program
paths that likely produced those messages, their technique achieved
improved verification latencies but still fell far short of being able
to keep pace with, e.g., highly interactive games.  Their approach
would not work for cryptographic protocols such as those we consider
here, since without substantial protocol-specific tuning, the
cryptographic protections would obscure information in messages on
which their technique depends for generating these hints.  In some
sense, our work in \secref{sec:multipass} can be viewed as providing a
method for iteratively stripping away these obfuscating effects of
cryptographic fields, with the bare minimum of manual configuration or
protocol-specific knowledge, and our work in \secref{sec:parallel}
then dampens the impact of generated hints that are ultimately
inaccurate.

Several other works have sought to verify the behavior of clients in
client-server protocols.  Most permit false rejections or acceptances
since they verify client behavior against an abstract (and so
imprecise) model of the client program
(e.g.,~\cite{giffin02:remote,guha09:ajax}), versus an actual client
program as we do here.  Others seek exact results as we do, but
accomplish this by modifying the client to send all inputs it
processes to the verifier, allowing the verifier to simply replay the
client on those inputs~\cite{vikram09:ripley}.  In our work, we verify
completely unchanged clients and introduce no additional messaging
overhead.  Proxies for inferring web-form parameter constraints when a
web form is served to a client, to detect parameter-tampering attacks
when the form values are returned~\cite{skrupsky13:tamperproof}, also
provide exact detection.  However, this work addresses only stateless
clients and does so without attention to cryptographically protected
traffic.  Our work permits stateful clients and specifically innovates
to overcome challenges associated with cryptographic protocols.

Also related to our goals are numerous works focused on verifying the
correctness of computations outsourced to an untrusted cloud.  Recent
works in this area, surveyed by Walfish and
Blumberg~\cite{walfish15:verifying}, employ advances in
probabilistically checkable proofs (e.g.,~\cite{ishai07:efficient})
and/or interactive proofs (e.g.,~\cite{goldwasser08:muggles}) to
permit a verifier to confirm (probabilistically) that an untrusted,
remote party performed the outsourced computation correctly, at a cost
to the verifier that is smaller than it performing the outsourced
computation itself.  Since we approach the problem from the opposite
viewpoint of a well-resourced verifier (e.g., running with the server
in a large cloud that the server owner trusts), our techniques do not
offer this last property.  However, ours requires no changes to the
party being verified (in our case, the client), whereas these other
works increase the computational cost for the party being verified (in
their case, the cloud) by orders of magnitude (e.g.,
see~\cite[\figrefstatic{5}]{walfish15:verifying}).  Another area of
focus in this domain has been reducing the privacy ramifications of
the additional information sent to the verifier to enable verification
(e.g.,~\cite{parno13:pinocchio}).  Since our technique does not
require changes to the messaging behavior of the application at all,
our technique does not suffer from such drawbacks.

More distantly related to our work is recent progress on reducing the
security of reference implementations of cryptographic protocols to
underlying cryptographic assumptions (e.g., miTLS, a reference
implementation of TLS in F\#~\cite{bhargavan13:tls}).  Whereas such
works prove specified properties of client and server implementations,
our work instead seeks to prove a property of the messages sent in an
interaction, the property being that these messages are consistent
with a specified client implementation.  As such, our techniques show
nothing about the intrinsic security of the client (or server)
implementation itself; nevertheless, they are helpful in detecting a
broad range of common exploit types in client-server protocols, as we
show here.  Our techniques also have the feature of being immediately
deployable to existing production protocol implementations.

\section{Background and Goals}
\label{sec:background}

A client-server protocol generates messages \msg{0}, \msg{1},
$\ldots$, some from the client and some sent by the server.  Our goal
is to construct a \textit{verifier} to validate the client behavior as
represented in the message sequence; the server is trusted.  We assume
that the client is single-threaded and that the message order reflects
the order in which the client sent or received those messages, though
neither of these assumptions is fundamental.  Our technique is not
dependent on a particular location for the verifier, though for the
purposes of this paper, we assume it is near the
server, acting as a passive network tap.\footnote{The verifier can
  optimistically assume that the order in which it observes the
  messages is the order in which the client sent or received them, and
  this assumption will often suffice to validate a legitimate client
  even if not strictly true, particularly when the client-server
  protocol operates in each direction independently (as in TLS).  In
  other cases, the verifier could in theory explore other orders when
  verification with the observed order fails.}

Borrowing terminology from prior work~\cite{cochran13:verification},
the task of the verifier is to determine whether there exists an
\textit{execution prefix} of the client that is \textit{consistent}
with the messages $\msg{0}, \msg{1}, \ldots$.  Specifically, an
execution prefix \execPrefix{} is a sequence of client instructions
that begins at the client entry point and follows valid branching
behavior in the client program.  We define \execPrefix{\msgNmbr} to be
\textit{consistent} with \msg{0}, \msg{1}, $\ldots$, \msg{\msgNmbr},
if the network \sendInstr and \recvInstr instructions\footnote{We
  abbreviate calls to POSIX \posixSend and
  \posixRecv system calls (or their functional equivalents) with the
  labels \sendInstr and \recvInstr.} in
\execPrefix{\msgNmbr} number $\msgNmbr+1$ and these network
instructions match \msg{0}, \msg{1}, $\ldots$, \msg{\msgNmbr} by
direction---i.e., if \msg{\msgIdx} is a client-to-server message
(respectively, server-to-client message), then the \msgIdx-th network
I/O instruction is a \sendInstr (respectively, \recvInstr)---and if
the branches taken in \execPrefix{\msgNmbr} were possible under the
assumption that $\msg{0}, \msg{1}, \ldots, \msg{\msgNmbr}$ were the
messages sent and received.  Consistency of \execPrefix{\msgNmbr} with
\msg{0}, \msg{1}, $\ldots$, \msg{\msgNmbr} requires that the
conjunction of all symbolic postconditions at \sendInstr instructions
along \execPrefix{\msgNmbr} be satisfiable, once concretized using
contents of messages \msg{0}, \msg{1}, $\ldots$, \msg{\msgNmbr} sent
and received on that path.

The verifier attempts to validate the sequence \msg{0}, \msg{1},
$\ldots$ incrementally, i.e., by verifying the sequence \msg{0},
\msg{1}, $\ldots$, \msg{\msgNmbr} starting from an execution prefix
\execPrefix{\msgNmbr-1} found to be consistent with \msg{0}, \msg{1},
$\ldots$, \msg{\msgNmbr-1}, and appending to it an \textit{execution
  fragment} that yields an execution prefix \execPrefix{\msgNmbr}
consistent with \msg{0}, \msg{1}, $\ldots$, \msg{\msgNmbr}.
Specifically, an \textit{execution fragment} is a nonempty sequence of
client instructions (i) beginning at the client entry point, a
\sendInstr, or a \recvInstr in the client software, (ii) ending at a
\sendInstr or \recvInstr, and (iii) having no intervening \sendInstr
or \recvInstr instructions.  If there is no execution fragment that
can be appended to \execPrefix{\msgNmbr-1} to produce a
\execPrefix{\msgNmbr} consistent with \msg{0}, \msg{1}, $\ldots$,
\msg{\msgNmbr}, then the search resumes by \textit{backtracking} to to
find another execution prefix \execPrefixAlt{\msgNmbr-1} consistent
with \msg{0}, \msg{1}, $\ldots$, \msg{\msgNmbr-1}, from which the
search resumes for an execution fragment to extend it to yield a
\execPrefixAlt{\msgNmbr} consistent with \msg{0}, \msg{1}, $\ldots$,
\msg{\msgNmbr}.  Only after all such attempts fail can the client
behavior be declared invalid, which may take substantial time.

Determining if a program can output a given value is only
semidecidable (recursively enumerable); i.e., while valid
client behavior can be declared as such in finite time, invalid
behavior cannot, in general.  Thus, an ``invalid'' declaration
usually comes by timeout on the verification
process.\footnote{Nevertheless, our tool declares our tested exploit
  traces as invalid within several seconds; see
  \secref{sec:evaluation:heartbleed}.}  However, our primary concern
in this paper is verifying the behavior of \textit{valid} clients
quickly.

\section{Parallel Client Verification}
\label{sec:parallel}

As discussed in the previous section, upon receipt of message
\msg{\msgNmbr}, the verifier attempts to find an execution fragment
with which to extend execution prefix \execPrefix{\msgNmbr-1}
(consistent with \msg{0}, $\ldots$, \msg{\msgNmbr-1}) to create an
execution prefix \execPrefix{\msgNmbr} that is consistent with
\msg{0}, $\ldots$, \msg{\msgNmbr}.
Doing so at a pace that keeps up with highly interactive applications
remains a challenge (e.g.,~\cite{cochran13:verification}).  We
observe, however, that multiple execution fragments can be explored
concurrently.  This permits multiple worker threads to symbolically
execute execution fragments simultaneously, while coordinating their
activities through data structures to ensure that they continue to
examine new fragments in priority order.  In this section we detail
the design of our tool to do so.

While concurrent exploration of execution fragments can improve the
performance of verification in any client-server application
(as we show in \appref{sec:parallel-demo}), there
are specific needs for this capability for verifying the cryptographic
protocols of primary interest in this paper.  For example, symbolic
execution tools such as the \klee tool on which we
build~\cite{cadar08:klee}, while being designed to work with program
variables whose \textit{values} are unknown (symbolic), nevertheless
require the \textit{sizes} of those variables to be fixed.  Upon
observing a message \msg{\msgIdx} that is encrypted, however, it may
not be possible to determine the size of the plaintext if padding is
added to the plaintext before encrypting (as in SSH and TLS 1.3)---and
in some cases, this might be exactly the reason that padding was
introduced (as in the case of TLS 1.3).  The ciphertext length does,
however, provide an upper bound on the plaintext length, and so
verification can proceed by considering each possible plaintext length
(up to the ciphertext length) individually.  Doing so sequentially
would result in a many-fold increase in verification cost, however.
Instead, by considering many plaintext lengths in parallel, the
increase in verification cost due to this ambiguity can be
substantially dampened.

\subsection{Algorithm overview}
\label{sec:parallel:overview}

We first define the data structures used by the algorithm. A state
\symState{} represents a snapshot of execution in the symbolic virtual
machine, including all constraints (path conditions) and memory
objects, which includes the contents (symbolic or concrete) of
registers, the stack and the heap.  We use \symState{}.\constraints to
represent the constraints accumulated during the execution to reach
\symState, and \symState{}.\nextInstruction to represent the next
instruction to be executed from \symState.
The verifier produces state \symState{\msgNmbr} by symbolically
executing the instruction sequence represented by execution prefix
\execPrefix{\msgNmbr}.

The algorithm builds and maintains a binary tree consisting of \Node
objects.  Each node \node has a field \node.\pathField to record a
path of instructions in the client; a field \node.\stateField that holds a
symbolic state; children fields $\node.\childField{0}$ and
$\node.\childField{1}$ that point to children nodes, and a field
$\node.\prevConstraintsField$ that will be described in
\secref{sec:multipass}.  The tree of nodes is rooted with a node \node
holding the state $\node.\stateField = \symState{\msgNmbr-1}$ and
$\node.\pathField = \execPrefix{\msgNmbr-1}$.  The two children of a
node \node in the tree extend \node.\pathField through the next
symbolic branch (i.e., branch instruction with a symbolic
condition). One child node holds a state with a constraint that
maintains that the branch condition implies false, and the other child
node's state holds a constraint that indicates that the branch
condition is true. The algorithm succeeds by finding a fragment with
which to extend \execPrefix{\msgNmbr-1} to yield \execPrefix{\msgNmbr}
if, upon extending a path, it encounters a network I/O instruction
that yields a state with constraints that do not contradict
\msg{\msgNmbr} being the network I/O instruction's message.

\begin{figure}[t]
\includegraphics[width=\columnwidth]{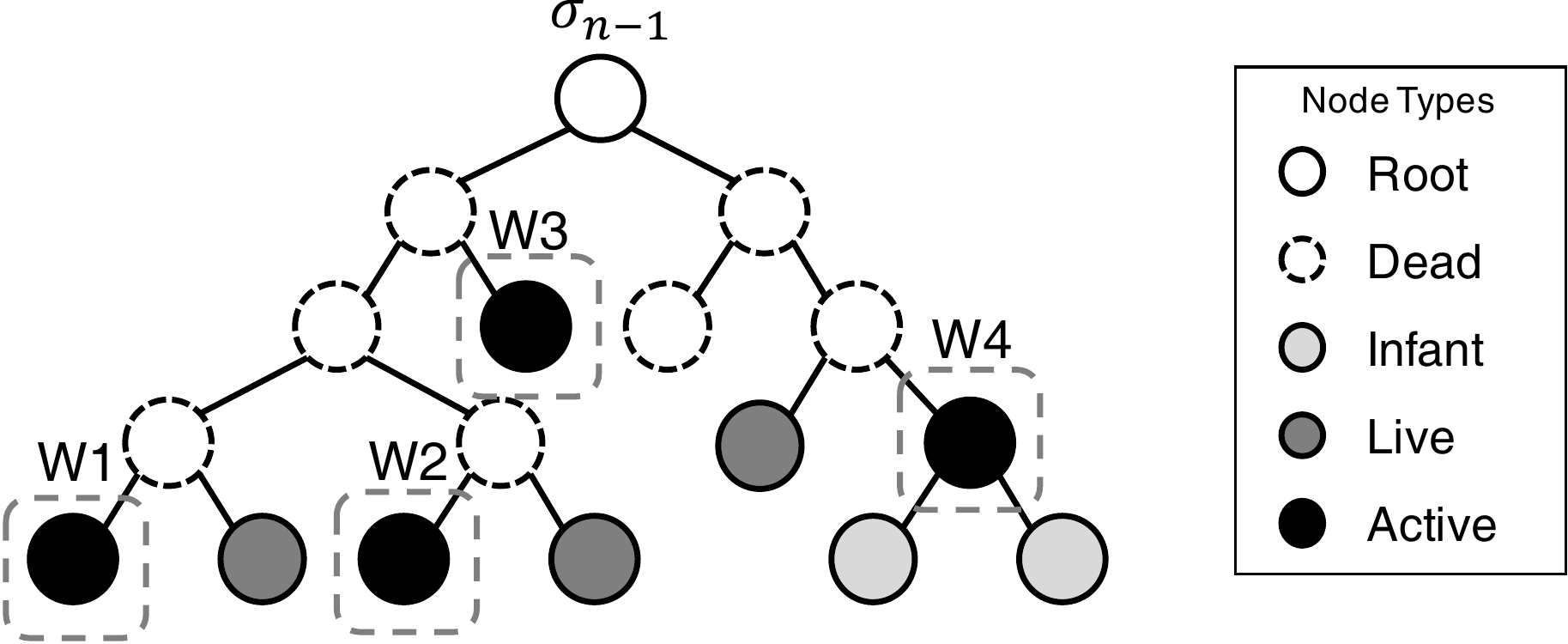}
\caption{Example node tree.
\label{fig:parallel:node}}
\end{figure}

The driving goal of our algorithm is to enable concurrent exploration
of multiple states in the node tree.  To this end, our parallel
verification algorithm uses multiple threads; it uses a single thread
to manage the node tree
and several worker threads, each
assigned to a single node in the node tree at a time.
\figref{fig:parallel:node} shows an example assignment of four workers
to multiple nodes in a node tree. In our design and experiments, the
number of worker threads \workerCount is a fixed parameter provided to
the verifier. Because the verification task is largely CPU-bound, in
our experience it is not beneficial to use more worker threads than
the number of logical CPU cores, and in some cases, fewer worker
threads than cores are necessary.

\subsection{Detailed algorithm description} 
\label{sec:parallel:verifydetails}

We will express our algorithm using standard multi-threading
primitives.  The keyword $\Spawn$
indicates the creation of
a child thread that will execute a named procedure until
completion, at which point the child thread will terminate. The
keyword $\Sync$ denotes that the parent procedure will not proceed to
the next statement until all spawned child threads have finished
execution.  In addition,
statements involving accesses to shared data structures will be
performed atomically (i.e., in a critical section), though for
readability we do not include explicit designation of the critical
section boundaries in our pseudocode.  Also for readability, our
pseudocode does not represent backtracking.

\begin{figure}[ht]
\begin{minipage}{\columnwidth}
\begin{algorithmic}[1]
\setalglineno{100}
\Procedure{\parallelVerifyAlg}{\execPrefix{\msgNmbr-1}, \symState{\msgNmbr-1}, 
                               \msg{\msgNmbr}}
  \label{fig:paralg:parallelVerifyAlg}
    \State $\rootNode \gets \makeNode(\execPrefix{\msgNmbr-1}, \symState{\msgNmbr-1}, \codeTrue)$
    \label{fig:paralg:initRoot}
    \State $\readyQ \gets \makeNodeQueue()$ \Comment{Nodes ready to execute}
      \label{fig:paralg:initReady}
    \State $\addedQ \gets \makeNodeQueue()$ \Comment{Nodes added by workers}
    \label{fig:paralg:initAdded}
    \label{fig:paralg:initWait}
    \State $\verifyFinished \gets \codeFalse$
    \Comment{Instructs all threads to halt}
    \label{fig:paralg:initFinished}
    \State $\resultNode \gets \makeNode(\bot, \bot, \bot)$
    \Comment Filled in on success
    \label{fig:paralg:initRslt}
    \State \Spawn \nodeScheduler{}(\rootNode, 
                                   \readyQ, \addedQ,
                                   \verifyFinished)
    %\Comment{Initialize Node Scheduler}
    \label{fig:paralg:spawnNodeScheduler}
    \For{1 \textbf{to} \workerCount}
    %\Comment{Spawn worker threads} 
    \State \Spawn \verifyWorker{}(\msg{\msgNmbr}, \rootNode, \readyQ, \addedQ,
    \verifyFinished, \resultNode)
      \label{fig:paralg:spawnVerifyWorker}
  \EndFor
  \State \Sync
    \Comment{Wait for all child threads} 
    \label{fig:paralg:sync}
  \State \Return \resultNode
    \Comment{Return result}
    \label{fig:paralg:returnValid}
\EndProcedure
\end{algorithmic}
\end{minipage}
\caption{Main procedure for parallel client verification.
\label{fig:paralg}}
\end{figure}

The algorithm for verifying a client-to-server message using
thread-level parallelism is shown in \figref{fig:paralg}. This
algorithm, denoted \parallelVerifyAlg, takes as input the execution
prefix \execPrefix{\msgNmbr-1} consistent with
$\msg{0},\ldots,\msg{\msgNmbr-1}$; the symbolic state
\symState{\msgNmbr-1} resulting from execution of
\execPrefix{\msgNmbr-1} from the client entry point on message trace
$\msg{0},\ldots,\msg{\msgNmbr-1}$; and the next message \msg{\msgNmbr}.
Its output is \resultNode, which holds the prefix
\execPrefix{\msgNmbr} and corresponding state \symState{\msgNmbr} in
$\resultNode.\pathField$ and $\resultNode.\stateField$, respectively,
if a prefix consistent with $\msg{0},\ldots,\msg{\msgNmbr}$ is found.
If the procedure returns with $\resultNode.\pathField =
\resultNode.\stateField = \bot$, then this indicates that there is no
execution prefix that can extend \execPrefix{\msgNmbr-1} to make
\execPrefix{\msgNmbr} that is consistent with \msg{0}, $\ldots$,
\msg{\msgNmbr}.  This will induce backtracking to search for another
\execPrefixAlt{\msgNmbr-1} that is consistent with \msg{0}, $\ldots$,
\msg{\msgNmbr-1}, which the verifier will then try to extend to find a
\execPrefixAlt{\msgNmbr} consistent with \msg{0}, $\ldots$,
\msg{\msgNmbr}.

The algorithm operates in a parent thread that spawns $\workerCount+1$
child threads; this includes
one thread to manage scheduling of nodes for execution via the
procedure \nodeScheduler (not shown) and \workerCount worker threads
to explore candidate execution fragments via the procedure
\verifyWorker (\figref{fig:paralg:verifyWorker}).

\nodeScheduler manages the selection of node states to execute next
and maintains the flow of nodes between worker threads.  It receives
as input two queues of nodes, a ``ready'' queue \readyQ and an
``added'' queue \addedQ. These queues are shared between the worker
threads and the \nodeScheduler thread. Worker threads pull nodes from
\readyQ and push new nodes onto \addedQ. As there is only one
scheduler thread and one or more worker threads producing and
consuming nodes from the queues \readyQ and \addedQ, \readyQ is a
single-producer-multi-consumer priority queue and \addedQ is a
multi-producer-single-consumer queue. The goal of \nodeScheduler is to
keep \addedQ empty and \readyQ full.  Nodes are in one of four
possible states, either actively being explored inside \verifyWorker,
stored in \readyQ, stored in \addedQ, or being prioritized by
\nodeScheduler.  A node at the front of \readyQ is the highest
priority node not currently being explored. The nodes in \addedQ are
child nodes that have been created by \verifyWorker threads that need
to be prioritized by \nodeScheduler and inserted into \readyQ.
\nodeScheduler continues executing until the boolean \verifyFinished
is set to true by some \verifyWorker thread.

Shown in \figref{fig:paralg:verifyWorker}, the procedure \verifyWorker
does the main work of client verification: stepping execution forward
in the state \newState of each node. In this figure, lines shaded gray
will be explained in \secref{sec:multipass} and can be ignored for now
(i.e., read \figref{fig:paralg:verifyWorker} as if these lines simply
do not exist).  Like \nodeScheduler, the procedure \verifyWorker runs
inside of a while loop until the value of \verifyFinished is no longer
equal to \codeFalse (\ref{fig:paralg:verifyWorkerWhile}). Recall that
the parent procedure \parallelVerifyAlg spawns multiple instances of
\verifyWorker. Whenever there is a node on the queue \readyQ, the
condition on \lineref{fig:paralg:ifDequeueReady} will be true and the
procedure calls \tryDequeue atomically. Note that even if $|\readyQ| =
1$, multiple instances of \verifyWorker may call \tryDequeue in
\ref{fig:paralg:dequeue}, but only one will return a node; the rest
will retrieve undefined ($\bot$) from \tryDequeue.

\begin{figure}[t!]
\begin{minipage}{\columnwidth}
\begin{algorithmic}[1]
\setalglineno{200}

\Procedure{\verifyWorker}{\msg{\msgNmbr}, \rootNode, \readyQ, \addedQ, \verifyFinished, \resultNode}
\While{$\neg\verifyFinished$}
  \label{fig:paralg:verifyWorkerWhile}
  \If{$|\readyQ| > 0$}
  \label{fig:paralg:ifDequeueReady}
    \State $\node \gets \tryDequeue(\readyQ)$
    \label{fig:paralg:dequeue}
    \If{$\node \neq \bot$}
      \label{fig:paralg:ifNode}
      \State $\newPath \gets \node.\pathField$ ; $\newState \gets \node.\stateField$ 
      \While{$\isNormal(\newState.\nextInstruction)$}
        \label{fig:paralg:whileSymEx}
        \State $\newPath \gets \newPath \parallel \langle \newState.\nextInstruction \rangle$
        \label{fig:paralg:extendPath}
        \State $\newState \gets \execStep(\newState)$
        \label{fig:paralg:execStep}
      \EndWhile
      
      \If{$\isIOInstruction(\newState.\nextInstruction)$}
      \label{fig:paralg:isIOInstruction}
      \If{$(\newState.\constraints\wedge\newState.\nextInstruction.\messageVar= \msg{\msgNmbr}) \not\Rightarrow \codeFalse$}
      \label{fig:paralg:consistent}
\graytrue
       \If{$(\newState.\constraints\wedge\newState.\nextInstruction.\messageVar \!=\! \msg{\msgNmbr}) \equiv \node.\prevConstraintsField$}
       \label{fig:paralg:nothingNew}
\grayfalse
            \State $\resultNode.\pathField \gets \newPath \parallel \langle \newState.\nextInstruction \rangle$
             \label{fig:paralg:successPath}
             \State $\resultNode.\stateField \gets$
             \Statex \hfill $[\execStep(\newState) \mid \newState.\nextInstruction.\messageVar \mapsto \msg{\msgNmbr}]$
             \label{fig:paralg:successState}
             \State $\verifyFinished \gets \codeTrue$
             \label{fig:paralg:returnSuccess}
             \Comment{Success!}
\graytrue
          \Else
            \State $\node \gets \clone(\rootNode)$
            \label{fig:paralg:rewind}
            \State $\node.\prevConstraintsField \gets \newState.\constraints\wedge\newState.\nextInstruction.\messageVar= \msg{\msgNmbr}$
            \label{fig:paralg:saveConstraints}
            \State $\enqueue(\addedQ, \node)$
            \label{fig:paralg:redo}
           \EndIf
        \EndIf
      \ElsIf{$\isProhibitive(\newState.\nextInstruction)$}
        \label{fig:paralg:isProhibitive}
        \State $\node.\pathField \gets \newPath \parallel \langle \newState.\nextInstruction \rangle$
        \State $\node.\stateField \gets \execSymbolicSkip(\newState,\node.\prevConstraintsField)$
        \label{fig:paralg:execSymbolicSkip}
        \State $\enqueue(\addedQ, \node)$
        \label{fig:paralg:endIsProhibitive}
\grayfalse
      \ElsIf{$\isSymbolicBranch(\newState.\nextInstruction)$}
        \label{fig:paralg:isSymbolicBranch}
        \State $\newPath \gets \newPath \parallel \langle \newState.\nextInstruction \rangle$
        \label{fig:paralg:falsePath}
        \State $\newStateAlt \gets \clone(\newState)$
        \State $\newStateAlt \gets [\execStep(\newStateAlt) \mid \newStateAlt.\nextInstruction.\condition \mapsto \codeFalse]$
        \label{fig:paralg:falseState}
        \If{$\newStateAlt.\constraints \not\Rightarrow \codeFalse$}
        \label{fig:paralg:checkFalse}
          \State $\node.\childField{0} \gets \makeNode(\newPath, \newStateAlt, \node.\prevConstraintsField)$\!\!
          \label{fig:paralg:makeChildFalse}
          \State $\enqueue(\addedQ, \node.\childField{0})$
          \label{fig:paralg:addFalse}
        \EndIf
        \State $\newState \gets [\execStep(\newState) \mid \newState.\nextInstruction.\condition \mapsto \codeTrue]$
        \label{fig:paralg:trueState}
        \If{$\newState.\constraints \not\Rightarrow \codeFalse$}
        \label{fig:paralg:checkTrue}
          \State $\node.\childField{1} \gets \makeNode(\newPath, \newState, \node.\prevConstraintsField)$
          \label{fig:paralg:makeChildTrue}
          \State $\enqueue(\addedQ, \node.\childField{1})$
          \label{fig:paralg:addTrue}
        \EndIf

      \EndIf
    \EndIf
  \EndIf

\EndWhile
\label{fig:paralg:mainWhileEnd}
\EndProcedure

\end{algorithmic}
\end{minipage}
\caption{\verifyWorker procedure.  Shaded lines will be explained in
  \secref{sec:multipass}.\label{fig:paralg:verifyWorker}}
\end{figure}

If \node is not undefined (\ref{fig:paralg:ifNode}), the algorithm
proceeds to execute the state $\node.\stateField$ and extend the
associated path $\node.\pathField$ up to either the next network
instruction (\sendInstr or \recvInstr) or the next symbolic branch (a
branch instruction that is conditioned on a symbolic variable). The
first case, stepping execution on a non-network / non-symbolic-branch
instruction \newState.\nextInstruction (here denoted
$\isNormal(\newState.\nextInstruction)$), executes in a while loop on
\linesref{fig:paralg:whileSymEx}{fig:paralg:execStep}.  The current
instruction is appended to the path and the procedure \execStep is
called, which symbolically executes the next instruction in state
\newState. These lines are where the majority of the computation work
is done by the verifier. The ability to concurrently step execution on
multiple states is where the largest performance benefits of
parallelization are achieved. Note that calls to \execStep may invoke
branch instructions, but these are non-symbolic branches.

In the second case, if the next instruction is \sendInstr or
\recvInstr and if the constraints $\newState.\constraints$ accumulated
so far with the symbolic state \newState do not contradict the
possibility that the network I/O message
$\newState.\nextInstruction.\messageVar$ in the next instruction
$\newState.\nextInstruction$ is \msg{\msgNmbr} (i.e.,
$(\newState.\constraints\wedge\newState.\nextInstruction.\messageVar =
\msg{\msgNmbr}) \not\Rightarrow \codeFalse$,
\lineref{fig:paralg:consistent}), then the algorithm has successfully
reached an execution prefix \execPrefix{\msgNmbr} consistent with \msg{0},
$\ldots$, \msg{\msgNmbr}.  The algorithm sets the
termination value ($\verifyFinished = \codeTrue$) and sets the return
values of the parent function on
\linesref{fig:paralg:successPath}{fig:paralg:successState}:
\resultNode.\pathField is set to the newly found execution prefix
\execPrefix{\msgNmbr} and \resultNode.\stateField is set to the state
that results from executing it, conditioned on the last message being
\msg{\msgNmbr} (denoted $[\execStep(\newState) \mid
  \newState.\nextInstruction.\messageVar \mapsto \msg{\msgNmbr}]$).
All other threads of execution now exit because $\verifyFinished =
\codeTrue$ and the parent procedure \parallelVerifyAlg will return
\resultNode.

In the final case, ($\isSymbolicBranch(\newState.\nextInstruction)$),
the algorithm is at a symbolic branch. Thus, the branch condition
contains symbolic variables and cannot be evaluated as true or false
in isolation. Using symbolic execution, the algorithm evaluates both
the true branch and the false branch by executing
\newState.\nextInstruction conditioned on the condition evaluating to
\codeFalse (denoted $[\execStep(\newStateAlt) \mid
  \newStateAlt.\nextInstruction.\condition \mapsto \codeFalse]$ in
\lineref{fig:paralg:falseState}) and conditioned on the branch
condition evaluating to \codeTrue (\ref{fig:paralg:trueState}). In
each case, the constraints of the resulting state are checked for
consistency (\ref{fig:paralg:checkFalse},
\ref{fig:paralg:checkTrue}), for example, using an SMT solver.
If either state is consistent, it is
atomically placed onto \addedQ (\ref{fig:paralg:addFalse},
\ref{fig:paralg:addTrue}).

\subsection{Algorithm summary}
\label{sec:parallel:summary}

Let us return to \figref{fig:parallel:node} from earlier, which
depicts a node tree rooted at \symState{\msgNmbr-1} during the
verification of \msg{\msgNmbr}. The node colored white with a solid
outline represents the \emph{root} node with state
\symState{\msgNmbr-1}. The nodes colored white with dashed outlines,
are the \emph{dead} nodes and represent intermediate states that no
longer exist. A node is dead when it does not reach a success
condition or exits the main \textbf{if} block of \verifyWorker
(starting on \lineref{fig:paralg:ifNode}) without generating any child
nodes.  Nodes colored black are the \emph{active} nodes and are
currently being explored by worker threads. Nodes colored dark gray
are being prioritized by \nodeScheduler and are still \emph{live}.  If
there are worker threads that are ready to process a node, the highest
priority live nodes are in \readyQ. Nodes colored light gray are the
\emph{infant} nodes and are in \addedQ. We can see that worker
\emph{W4} recently hit a symbolic branch condition and created two
infant nodes which were added to \addedQ. The other workers are likely
executing \linesref{fig:paralg:whileSymEx}{fig:paralg:execStep}.

While we are mainly concerned with cryptographic protocols, in
\appref{sec:parallel-demo} we show that our parallel algorithm can be
highly effective in improving the speed of verifying clients in other
distributed applications.

\section{Multipass Client Verification}
\label{sec:multipass}

As shown in \appref{sec:parallel-demo}, concurrent exploration
of execution fragments can be highly beneficial to the speed of
validating legitimate client behavior in non-cryptographic protocols.
For verifying a cryptographic client, concurrent exploration of
execution fragments
can be similarly beneficial, as we will show in
\secref{sec:eval:padding}.  Nevertheless, there remain challenges to
verifying cryptographic clients that no reasonable amount of
parallelization can overcome, since doing so would be tantamount to
breaking some of the underlying cryptographic primitives themselves.
In this section, we introduce a strategy for client verification that
can overcome these hurdles for practical protocols such as TLS.

The most obvious challenge is encrypted messages.  To make
sense of these messages, the verifier will need to be
given the symmetric session key under which they are encrypted.
Fortunately, existing implementations of, e.g., \openssl servers,
enable logging session keys to support analysis of network captures,
and so we rely on such facilities to provide the session key to the
verifier.  Given this, it is theoretically straightforward to reverse
the encryption on a client-to-server message mid-session---just as the
server can---but that capability does surprisingly little to itself
aid the verification of the client's behavior.  Indeed,
state-of-the-art servers routinely fail to detect problems with the
message sequence received from a client, as demonstrated by numerous
such CVEs over the past two years in all major TLS
implementations~\cite{leyden14:tlspwned}.

We therefore continue with our strategy of incrementally building an
execution prefix \execPrefix{} in the client software as each message
is received by the verifier to validate the client's behavior.  The
verifier injects the logged session key into the execution prefix at
the point where the key would first be generated by the client.
Still, however, the number of execution fragments that need to be
explored in cryptographic protocol client implementations is far too
large to overcome by concurrent exploration alone, when other inputs
to cryptographic algorithms can be symbolic.  Some of these (e.g., a
message plaintext, once decrypted) could be injected by the verifier
like the session key is, but in our experience, configuring where to
inject what values would require substantially greater
client-implementation-specific knowledge and bookkeeping than
injecting just the session key does.  This is in part due to the many
layers in which cryptographic mechanisms are applied in modern
protocols; e.g., in the TLS handshake, multiple messages are hashed to
form the plaintext of another message, which is subsequently encrypted
and authenticated.  Even worse, other values, such as a client's
ephemeral Diffie-Hellman key, will never become available to a
verifier (or server) and so cannot be injected into an execution
prefix \execPrefix{}.

These observations motivate a design whereby the verifier is permitted
to skip specified functions that would simply be too expensive to
execute with symbolic inputs.  Specifying such \textit{prohibitive
  functions} need not require substantial
client-implementation-specific or even protocol-specific knowledge; in
our experience with TLS, for example, it suffices to specify basic
cryptographic primitives such as modular exponentiation, block
ciphers, and hash functions as prohibitive.  Once specified as
prohibitive, the function is skipped by the verifier if any of its
inputs are symbolic, producing a symbolic result instead.  Once
reconciled with the message sequence \msg{0}, $\ldots$, \msg{\msgNmbr}
under consideration, however, the verifier can solve for some values
that it was previously forced to keep symbolic, after which it can go
back and verify function computations (concretely) it had previously
skipped.  Once additional passes yield no new information, the
verifier outputs any unverified function computations (e.g., ones
based on the client's ephemeral Diffie-Hellman key) as assumptions on
which the verification rests.
Only if one of these assumptions is not true will our verifier
erroneously accept this message trace.  As we will see, however,
these remaining assumptions for a protocol like TLS are minimal.

\subsection{User configuration}
\label{sec:multipass:config}

As mentioned previously, our algorithm requires the specification of
prohibitive functions.  A prohibitive function is required to have no
side effects other than altering its own parameters (or parameter
buffers if passed by reference) and producing a return value; given
the same inputs, it must produce the same results; and it must be
possible to compute the sizes of all output buffers as a function of
the sizes of the input buffers.  A function should be specified as
prohibitive if executing it on symbolic inputs induces a large number
of symbolic states, due to branching that depends on
input values.  For example, a physics engine might contain signal
processing functions that should be marked prohibitive.

In our case studies, the prohibitive functions are cryptographic
functions such as the AES block cipher or SHA-256.  We stress,
however, that the user need not know how these primitives are composed
into a protocol.  We illustrate this in \appref{sec:tls-setup}, where
we show the user configuration needed for verifying the \openssl
client, including the specification of the prohibitive functions.

Specifying prohibitive functions generalizes the normal
procedure used by symbolic execution to inject symbolic inputs into
the program.  The user normally designates ``user input''
functions (such as \texttt{getchar}) as symbolic, so that each one is
essentially replaced with a function that always returns a symbolic,
unconstrained value of the appropriate size.  The random number
generators, client-side inputs (i.e., \stdin), and functions that
return the current time are typically so designated.  The user
configuration for prohibitive functions simply extends this mechanism
so that some of these functions do not always return symbolic outputs,
but return concrete outputs when their inputs are fully concrete.

\subsection{Algorithm description}
\label{sec:multipass:algo}

The multipass verification algorithm involves changes to the
\verifyWorker procedure in \figref{fig:paralg:verifyWorker},
specifically the insertion of the shaded lines.  Whenever
\newState.\nextInstruction is a call to a prohibitive function, it is
treated separately
(\linesref{fig:paralg:isProhibitive}{fig:paralg:endIsProhibitive}),
using the \execSymbolicSkip function
(\ref{fig:paralg:execSymbolicSkip}).  (To accomplish this, \isNormal
in \lineref{fig:paralg:whileSymEx} now returns \codeFalse not only for
any network instruction or symbolic branch, but also for any call to a
prohibitive function.) If \execSymbolicSkip receives a call
\newState.\nextInstruction to a prohibitive function with any symbolic
input buffers, it replaces the call with an operation producing fully
symbolic output buffers of the appropriate size.  However, if the
constraints saved in \node.\prevConstraintsField allow the concrete
input buffer values to be inferred, then \execSymbolicSkip instead
performs the call \newState.\nextInstruction on the now-concrete input
buffers.

Prior to the execution path reaching a network instruction, when a
call \newState.\nextInstruction to a prohibitive function is
encountered, \node.\prevConstraintsField is simply \codeTrue as
initialized (see the third argument to \makeNode in
\lineref{fig:paralg:initRoot} of \figref{fig:paralg}, as well as in
\ref{fig:paralg:makeChildFalse} and \ref{fig:paralg:makeChildTrue}),
permitting no additional inferences about the values of input buffers
to \newState.\nextInstruction.  After a network instruction is reached
and \msg{\msgNmbr} is reconciled with the constraints
\newState.\constraints accumulated along the path so far
(\ref{fig:paralg:consistent}), the path constraints
\newState.\constraints and the new constraint
$\newState.\nextInstruction.\messageVar = \msg{\msgNmbr}$ are saved in
\node.\prevConstraintsField (\ref{fig:paralg:saveConstraints}).  The
execution path is then replayed from the root of the binary tree
(i.e., beginning from \execPrefix{\msgNmbr-1}, see
\ref{fig:paralg:rewind}).  This process repeats until an execution
occurs in which nothing new is learned (i.e.,
$(\newState.\constraints\wedge\newState.\nextInstruction.\messageVar =
\msg{\msgNmbr}) \equiv \node.\prevConstraintsField$, in
\ref{fig:paralg:nothingNew}), at which point \verifyWorker returns as
before.

\subsection{Detailed walk-through}
\label{sec:multipass:detailedwalk}

In \secref{sec:multipass:algo}, we concisely summarized the multipass
algorithm as defined by the shaded lines
in \figref{fig:paralg:verifyWorker}. We now provide a detailed
walk-through of the algorithm on a segment of C code that includes both
an encryption operation and a network operation
(\figref{fig:clientexample:code}), starting from entry point and
following the verifier to termination.

%%%%%%%%%%%%%%%%%%%%%%%%%%%%%%%%%%%%%%%%%%%%%%%%%%%%%%%%%%%%%%%%%%%%%%%%

\newbox\mybox
\begin{lrbox}{\mybox}
\lstset{numbers=none,basicstyle=\small\ttfamily}
\begin{lstlisting}
void Client(int x, int y, int iv) {
	int p = x*y;
	if (x % 9 == 0) {
		if (y & 1 == 1) {
			int s = AES(iv);
			int c = p ^ s;
			SEND(iv, c);
		}
	}
}
\end{lstlisting}
\end{lrbox}

\begin{figure}[t]
\subfloat[][Example client code]{
\label{fig:clientexample:code}
\usebox\mybox
}\\
\subfloat[][Pass one]{
  \label{fig:clientexample:passone}
  \includegraphics[width=\columnwidth]{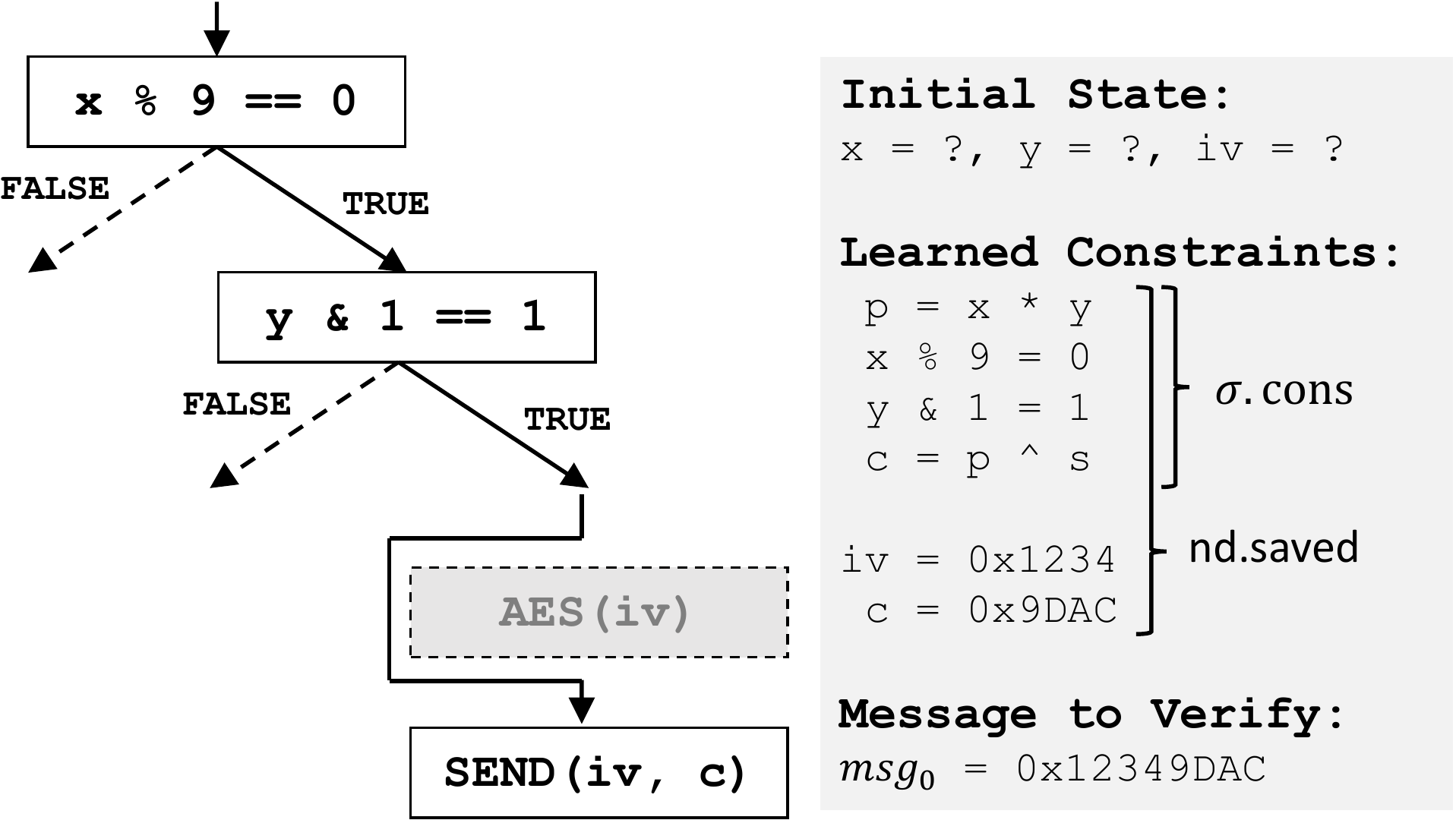}
}\\
\subfloat[][Pass two]{
  \label{fig:clientexample:passtwo}
  \includegraphics[width=\columnwidth]{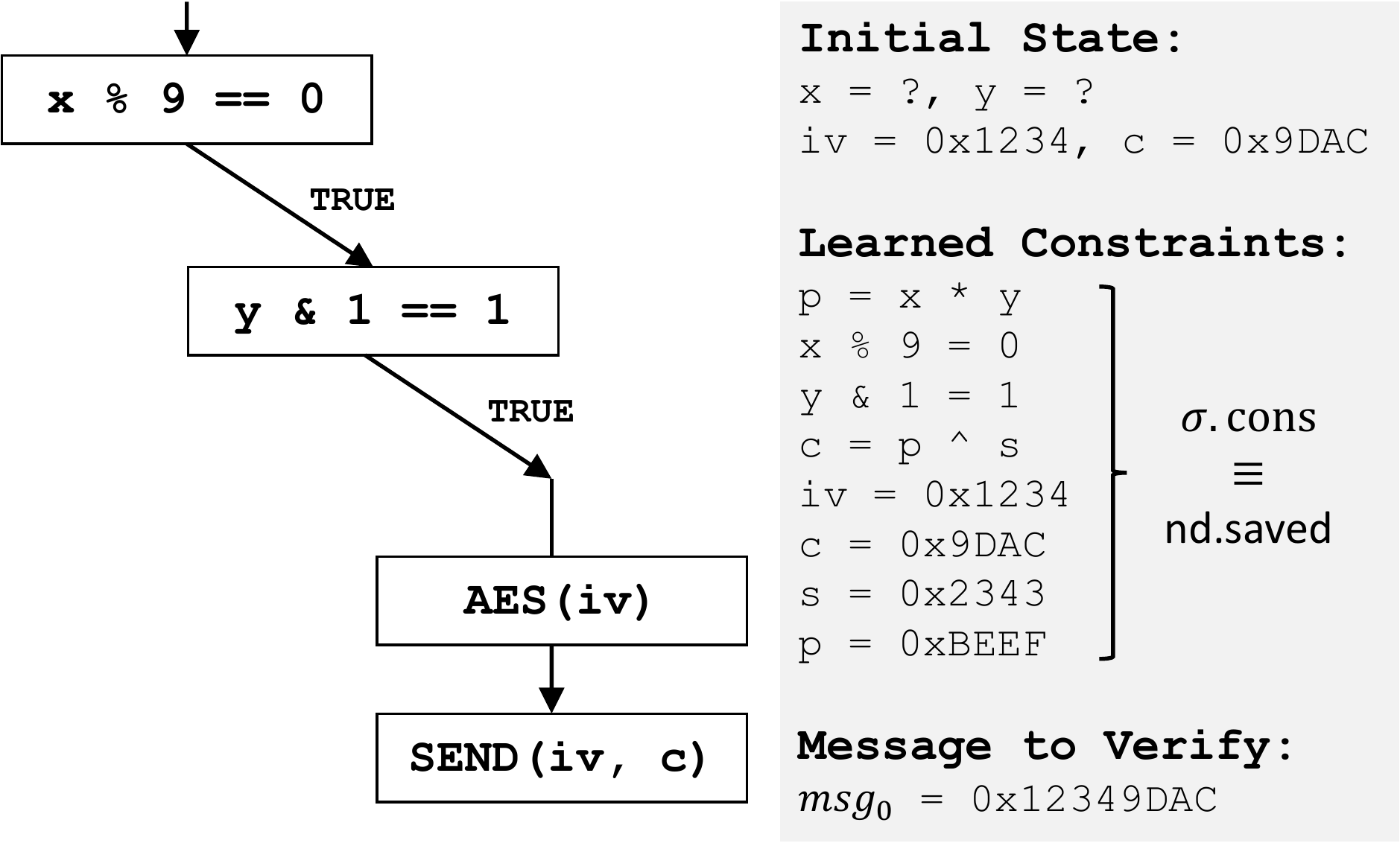}
}
\caption{Example of multipass verification on a simple client.
\label{fig:clientexample}}
\end{figure}

This client multiplies two of its inputs \texttt{x} and \texttt{y},
encrypts it using a third input \texttt{iv} as an initialization
vector, and sends both \texttt{iv} and the encrypted value to the
server.  Our tool begins with a node initialized to the client entry
point and attempts to verify (by spawning one or more threads that
execute \verifyWorker) that there exist inputs \texttt{x}, \texttt{y},
and \texttt{iv} that would produce the output message $\msg{0}
= \verb|0x12349DAC|$ that was observed over the network.

The instance of
\verifyWorker that first reaches the \sendInstr has, by that time,
accumulated constraints \newState.\constraints as specified in
\figref{fig:clientexample:passone}.  Note, however, that it has no
constraints relating \texttt{s} (the output of \texttt{AES(iv)}) and
\texttt{iv}, since \texttt{AES} was designated as prohibitive and
skipped (since \texttt{iv} is symbolic).  After reconciling these
constraints with the message $\msg{0} = \verb|0x12349DAC|$, the
verifier records \node.\prevConstraintsField.

The verifier then re-executes from \rootNode
(\figref{fig:clientexample:passtwo}), although since it now knows
\verb|iv = 0x1234|, this time it does not skip \texttt{AES}.  It thus
computes a concrete output \verb|s = 0x2343| and the constraint
\verb|0x9DAC = p ^ 0x2343|, i.e., \verb|p = 0xBEEF|.  After this
second pass, the constraints in \node.\prevConstraintsField are still
satisfiable (e.g., \verb|x = 0x9|, \verb|y = 0x1537|).  However, the
third pass (not shown) reveals no more information, so \verifyWorker
returns the corresponding execution prefix and state at the end of the
third pass.

\subsection{TLS example}
\label{sec:multipass:tls}

\begin{figure*}[ht]
\centering
\subfloat[][First pass, execution]{
  \label{fig:multipass1a}
  \includegraphics[width=\columnwidth]{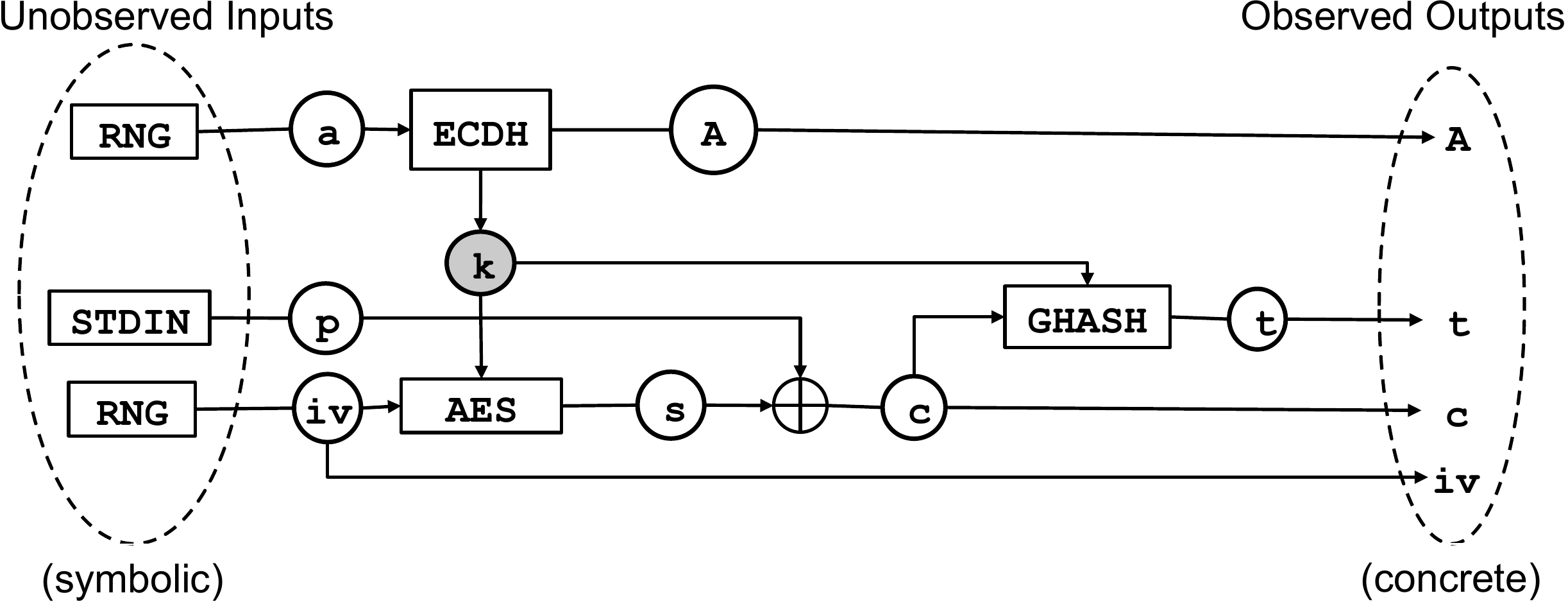}
}
\subfloat[][First pass, reconciliation]{
  \label{fig:multipass1b}
  \includegraphics[width=\columnwidth]{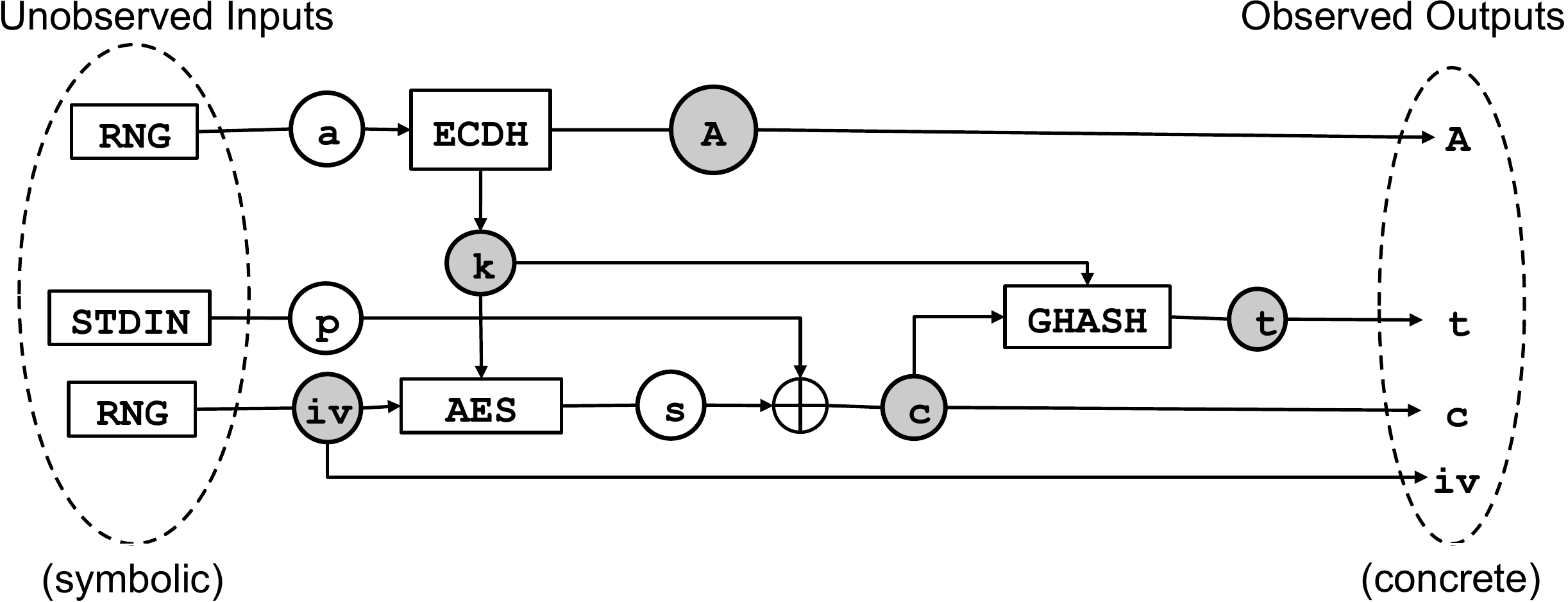}
}\\
\subfloat[][Second pass, execution]{
  \label{fig:multipass2a}
  \includegraphics[width=\columnwidth]{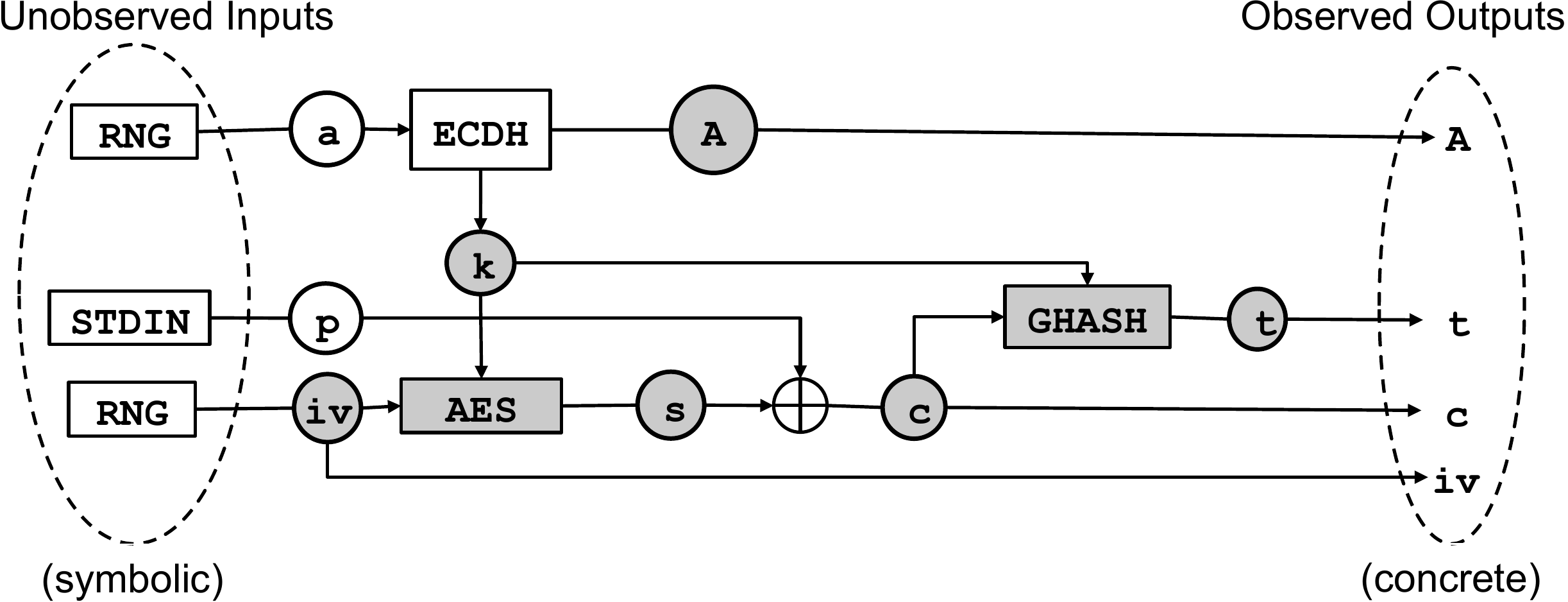}
}
\subfloat[][Second pass, reconciliation]{
  \label{fig:multipass2b}
  \includegraphics[width=\columnwidth]{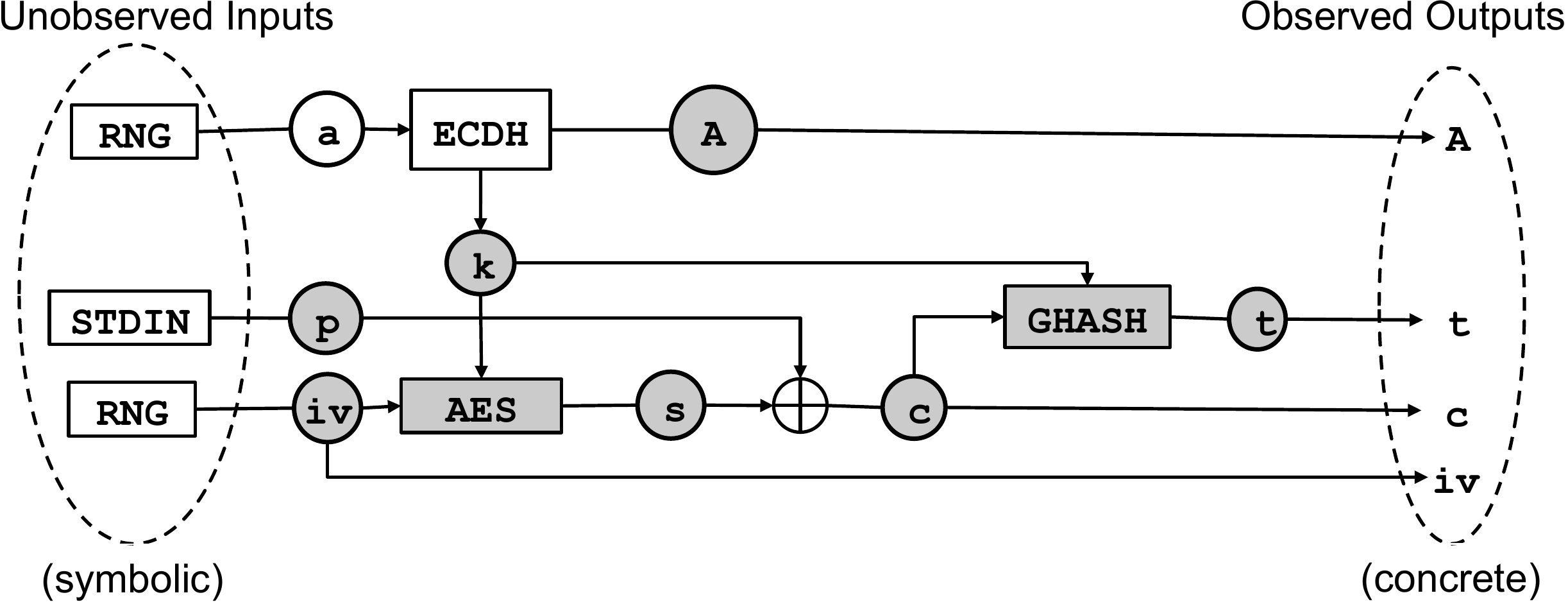}
}
\caption{Multipass algorithm on a TLS client implementing an
  abstracted subset of AES-GCM. Rectangular blocks are prohibitive
  functions; circles are variables.  Shaded nodes are concrete values
  or functions executed with concrete inputs. Unshaded nodes are
  symbolic values or skipped functions.  In \figref{fig:multipass1b}
  and \figref{fig:multipass2b}, some values become concrete when
  \newState.\constraints is reconciled with
  $\newState.\nextInstruction.\messageVar = \msg{\msgNmbr}$ in
  \lineref{fig:paralg:consistent}.}
\label{fig:multipass}
\end{figure*}

We illustrate the behavior of the multipass algorithm on TLS.
\figref{fig:multipass} shows an
abstracted subset of a TLS client implementation of AES-GCM, running
on a single block of plaintext input.  For clarity, the example omits
details such as the implicit nonce, the server ECDH parameters, the
generation of the four symmetric keys, and subsumes the tag
computation into the GHASH function. But in all features shown, this
walkthrough closely exemplifies the multi-pass verification of a
real-world TLS client.

In \figref{fig:multipass}, the outputs observed by the verifier are
the client Diffie-Hellman parameter \texttt{A}, the initialization
vector \texttt{iv}, the ciphertext \texttt{c}, and the AES-GCM tag
\texttt{t}.  The unobserved inputs are the Diffie-Hellman private
exponent \texttt{a}, the initialization vector \texttt{iv}, and the
plaintext \texttt{p}.  We do assume access to the AES symmetric key
\texttt{k}.  Since the client verification is being performed on the
server end of the connection, we can use server state, including the
symmetric key.  The verifier decides whether the observed outputs are
legal, given knowledge of the program but not its inputs.

In the first pass of symbolic execution (\figref{fig:multipass1a}),
even with knowledge of the AES symmetric key \texttt{k}, all
prohibitive functions (ECDH, AES, GHASH) have at least one symbolic
input.  So, \execSymbolicSkip skips them and produces unconstrained
symbolic output for each.  After the first execution pass
(\figref{fig:multipass1b}), the verifier encounters the observed
client outputs.  Reconciling them with the accumulated constraints
\newState.\constraints yields concrete values for \texttt{A},
\texttt{t}, \texttt{c}, and \texttt{iv}, but not the other variables.

The verifier then begins the second pass of symbolic execution
(\figref{fig:multipass2a}).  At this point, AES and GHASH both have
concrete inputs, and therefore can be executed concretely.  Note that
the concrete execution of AES yields a concrete value for \texttt{s},
which was not previously known.  At the end of the second execution
pass (\figref{fig:multipass2b}), the verifier implicitly uses the new
knowledge of \texttt{s} to check that there exists a \texttt{p}, the
unobserved plaintext value, that satisfies the constraints imposed by
observed output.  Further passes beyond this point yield no additional
information, as no further symbolic inputs to prohibitive functions
can be concretized.

Note that the value of \texttt{a}, the client Diffie-Hellman private
exponent, is never computed.  The verifier thus yields as output an
additional assumption that there exists an \texttt{a} such that
$\texttt{ECDH(a)}$ yields values \texttt{A} and \texttt{k}.  As such,
we do not detect invalid curve attacks~\cite{jager:2015:invalid}, for
example; we discuss practical mitigations for this in
\secref{sec:discussion:limitations}.

Perhaps remarkably, no decryption mechanism is explicitly provided to
the verifier.  The multipass mechanism automatically recovers the
plaintext for stream ciphers and counter-mode block ciphers. For other
modes such as CBC, the user could provide inverse functions via an
extension described in \appref{sec:constraint-gen}.

\section{Implementation} 
\label{sec:implementation}

We have designed and implemented a prototype of our client
verification technique. Our implementation is built upon a modified
version of \klee~\cite{cadar08:klee} and employs optimizations used in
previous work~\cite{cochran13:verification} as well as several new
performance improvements.  At a high level, \klee serves as an
interpreter of LLVM assembly instructions.  When an instruction has
symbolic operands, the operation is stored as a symbolic expression;
otherwise the operation is interpreted concretely. Note that even
interpreting instructions concretely induces a significant performance
penalty compared to native execution.  We reduce this cost by
leveraging the information provided in the user configuration of
prohibitive functions to mix in native execution of calls to
prohibitive functions with concrete inputs.
The only additional requirement for this optimization is to provide
the verifier with a native shared object that exports implementations
of the prohibitive functions.

Symbolic execution of the client requires interaction with the
environment through library and system calls. Our implementation
inherits and extends several mechanisms from \klee for interacting
with the environment. We use a combination of the POSIX model provided
by \klee, and an expanded POSIX environment model from another system,
\cloudnine~\cite{bucur11:cloud9}.  We also make two key extensions to
this model to support network instructions and prohibitive functions.
First, during symbolic execution of the client, each \sendInstr or
\recvInstr is is intercepted and the next message in the trace under
verification for a given symbolic state is processed as described in
\secref{sec:parallel}.  POSIX network calls are thus modeled to be
consistent with the message trace we are verifying. Prohibitive
functions are supported through an extension to the \klee runtime
model. The steps a user takes to add a prohibitive function to a given
client program are straightforward. First, the user adds the function
signature to a specified file in the \klee runtime model using an API
provided by the verifier. Next, the user defines the input parameters
and their size, as well as the output parameters and size (as a
function of the input size). The model is compiled and linked with the
client program to prepare for verification. When a prohibitive
function is reached during verification, the verifier uses the
user-provided annotations to identify if any input memory is symbolic
and, if so, the function is ``skipped'' and the outputs are
initialized and marked as symbolic.  If all inputs are concrete, then
the underlying implementation is executed.

The parallel client verification algorithm (\secref{sec:parallel})
requires a mechanism to concurrently execute symbolic states. We
extended \klee to support multi-threaded operation. Our changes
did not significantly alter the overall architecture of \klee, but
several submodules required modification to support concurrent
execution, including symbolic state ``searchers'', symbolic memory
management, and constraint caching and solving.

To our knowledge, no current symbolic execution engines support
parallel execution in a single process via multiple threads.  While
other efforts have demonstrated the feasibility and performance
benefits of parallelized symbolic execution engines
(\cite{bucur11:cloud9, siddiqui10:parsym, staats10:pse}), these
approaches differ from ours by dividing the symbolic execution work
across multiple processes or hosts instead of across multiple threads.
A multi-threaded symbolic execution engine can leverage opportunities
to identify duplicate states, utilize state merging, and utilize
shared constraint-solving caches. Furthermore, our architecture does
not incur costs due to the latency of communication between multiple
hosts; our application of symbolic execution to client verification
requires high speed context switching between states.  Additionally
our state selection can achieve efficiencies with global knowledge of
the progress of each execution path. Finally, our verifier is designed
to solve SMT queries concurrently with multiple instantiations of an
SMT solver, in our case STP~\cite{ganesh07:stp}.

\section{Evaluation}
\label{sec:evaluation}

In this section we evaluate our implementation of the algorithms in
\secsref{sec:parallel}{sec:multipass}.
First, we run a single-worker verifier
against two attacks on \openssl that represent different classes of
client misbehavior.  Second, we load test a single-worker verifier on
a typical TLS 1.2 payload, i.e., the traffic generated by a \gmail
session.  Third, we increase the verification complexity to
demonstrate scalability to more complex protocols with larger client
state spaces, which we overcome using multiple workers.  We do this by
simulating verification of a TLS 1.3 draft~\cite{tls13} feature that
permits arbitrary random padding in every packet.  The \openssl
configuration options we used are listed in \appref{sec:tls-setup}.
All experiments were run on a system with $256\gbytes$ of RAM and
$3.2\ghertz$ processor cores.

Our primary measure of performance is verification \textit{lag}.  To
define lag, let the verification \textit{cost} of a message
\msg{\msgNmbr}, denoted \cost{\msgNmbr}, be the wall-clock time that
the verifier spends to conclude if \msg{\msgNmbr} is valid, beginning
from the execution prefix \execPrefix{\msgNmbr-1} produced from the
successful verification of $\msg{0}, \ldots, \msg{\msgNmbr-1}$.  Since
the verifier is compute-bound, \cost{\msgNmbr} is roughly the CPU time
that it spends to produce \execPrefix{\msgNmbr} from
\execPrefix{\msgNmbr-1}.\footnote{In case of backtracking---i.e., if
  at any point an alternate prefix \execPrefixAlt{\msgNmbr-1} must be
  produced---then the time needed to advance
  \execPrefixAlt{\msgNmbr-1} to an \execPrefixAlt{\msgNmbr} that is
  consistent with all of $\msg{0},\ldots,\msg{\msgNmbr}$ is also
  accumulated into \cost{\msgNmbr}.}  The
\textit{completion time} for \msg{\msgNmbr} is then defined
inductively as follows:
\begin{align*}
\completion{0} & = \cost{0} \\
\completion{\msgNmbr} & = \max\{\arrival{\msgNmbr},\completion{\msgNmbr-1}\} + \cost{\msgNmbr}
\end{align*}
where \arrival{\msgNmbr} is the wall-clock time when \msg{\msgNmbr}
arrived at the verifier.  Since the verification of \msg{\msgNmbr}
cannot begin until after both (i) it is received at the verifier (at
time \arrival{\msgNmbr}) and (ii) the previous messages
$\msg{0},\ldots,\msg{\msgNmbr-1}$ have completed verification (at time
\completion{\msgNmbr-1}), \completion{\msgNmbr} is calculated as the
cost \cost{\msgNmbr} incurred after both (i) and (ii) are met.
Finally, the \textit{lag} of \msg{\msgNmbr} is $\lag{\msgNmbr} =
\completion{\msgNmbr} - \arrival{\msgNmbr}$, which is our primary
measure of performance.

\subsection{Heartbleed and CVE-2015-0205 detection}
\label{sec:evaluation:heartbleed}

We first evaluate our client verifier against two attacks on \openssl,
which are meant to be illustrative of different classes of vulnerabilities
that we can verify: those related to tampering with the client software
to produce messages that a client could not have produced (CVE-2014-0160 \heartbleed)
and a message sequence that, while correctly formatted, is impossible
given a valid client state machine (CVE-2015-0205).  

An \openssl 1.0.1f \sserver was instantiated with standard settings, and
an \openssl \sclient was modified so that it would establish
a TLS connection and send a single Heartbleed exploit packet.  This
packet had a modified length field, and when received by an \openssl
1.0.1f \sserver, caused the server to disclose sensitive
information from memory.

When client verification was applied to a normal client (which sends a
normal Heartbeat packet), the verification lag for the (last message
of the) connection was 1.7\secs.  When it was applied to a client that
sends a Heartbleed exploit, the validation process was unable to find
an explanation for the contents of the packet and rejected the packet
after exhausting all search paths, with a lag for the Heartbleed
packet of 6.9\secs.

Unlike \heartbleed, CVE-2015-0205 contains only correctly formatted 
messages.  In the certificate exchange, a good client will send a
DH certificate (used to generate a pre-master secret), followed by
a 0-length ClientKeyExchange message.  A malicious client will
send a certificate, followed by a ClientKeyExchange message containing a
DH parameter.  The server will authenticate the certificate, but prefer
the second message's DH parameter, allowing a malicious client to
impersonate anyone whose public certificate it has obtained.
We introduced this vulnerability to an \openssl 1.0.1d \sserver which 
was instantiated with standard settings, and an \openssl \sclient was modified 
to send a ClientKeyExchange containing a DH parameter.  The server 
authenticated the certificate but preferred the second DH parameter.

The verification lag for the good connection was 1.3\secs, and the
verifier rejected an attempted attack after a lag of 2.4\secs.  This
exploit illustrates the power of our technique: we not only verify
whether each message is possible in isolation, but also in the context
of all previous messages.

Since the tool verifies \textit{valid} client behavior, no
attack-specific configuration was required.  We do not require any
foreknowledge of the exploit and anticipate correct detection of other
exploits requiring client tampering.

\subsection{Performance evaluation: Typical TLS load}
\label{sec:evaluation:gmail}

The \gmail performance test was designed to measure the lag that
would result from running a single-worker verifier against typical real-world
TLS traffic.  The data set was a \texttt{tcpdump} capture of a
three-minute \gmail session conducted in Firefox, and consisted of 21
concurrent, independent TLS sessions, totaling 3.8\mbytes of network
data.  This \gmail session was performed in the context of one of the
authors' email accounts and included both receiving emails and sending
emails with attachments.

The verification objective of this test was to verify the TLS layer of
a network connection, but not the application layer above it, such as
the browser logic and \gmail web application.  To simulate the
client-server configuration without access to \gmail servers and
private keys, we used the packet sizes and timings from the \gmail
\texttt{tcpdump} to generate 21 equivalent sessions using \openssl
\sclient and \sserver, such that the amount of traffic sent in each
direction at any point in time matches identically with that of the
original \gmail capture.  Since \sclient implements a few diagnostic
features in addition to TLS (but no application layer), verifying
\sclient against these 21 sessions provides a conservative evaluation
of the time required to verify the pure TLS layer.

\begin{figure}[t]
  %\centering
\hspace{-0.75em}
\subfloat[][Client-to-server volumes]{
\label{fig:bw:c2s}
\includegraphics[width=0.5\columnwidth]{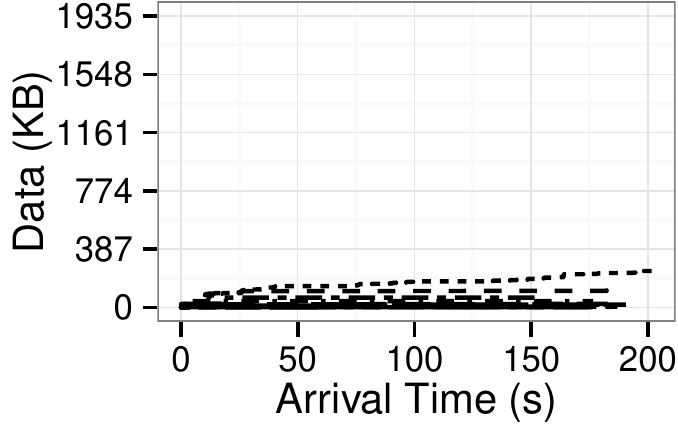}
}
\subfloat[][Server-to-client volumes]{
\label{fig:bw:s2c}
\includegraphics[width=0.5\columnwidth]{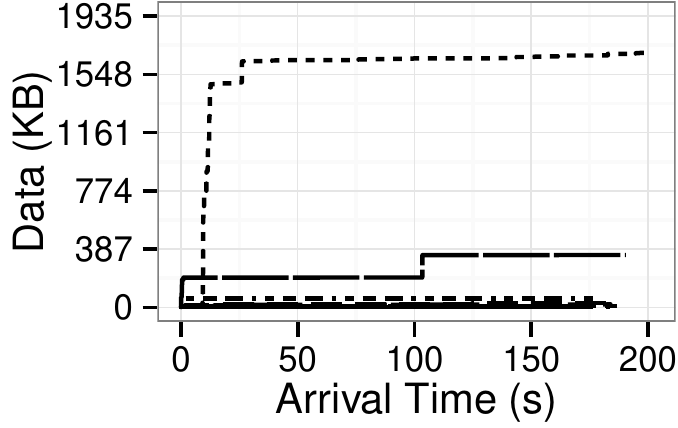}
}
\caption{Cumulative data transferred for a \gmail session consisting of
  21 TLS connections.  
  \figref{fig:bw:c2s} shows the volumes transferred in the
  client-to-server direction, with one line per TLS session.
  \figref{fig:bw:s2c} shows the
  volumes transferred in the server-to-client direction.}
\label{fig:bw}
\end{figure}

\figref{fig:bw} shows the bandwidth characteristics of these 21 TLS
connections.  As can be seen, one of the 21 TLS sessions was
responsible for the vast majority of the data transferred, and almost
all of the data it carried was carried from the server to the client
(\figref{fig:bw:s2c}); presumably this was a bulk-transfer connection
that was involved in prefetching, attachment uploading, or other
latency-insensitive tasks.  The other 20 TLS sessions were utilized
comparatively lightly and presumably involved in more
latency-sensitive activities.

\begin{figure}[t]
  %\centering
\hspace{-0.75em}
\subfloat[][All 21 TLS sessions]{
\label{fig:lag:all}
\includegraphics[width=0.5\columnwidth]{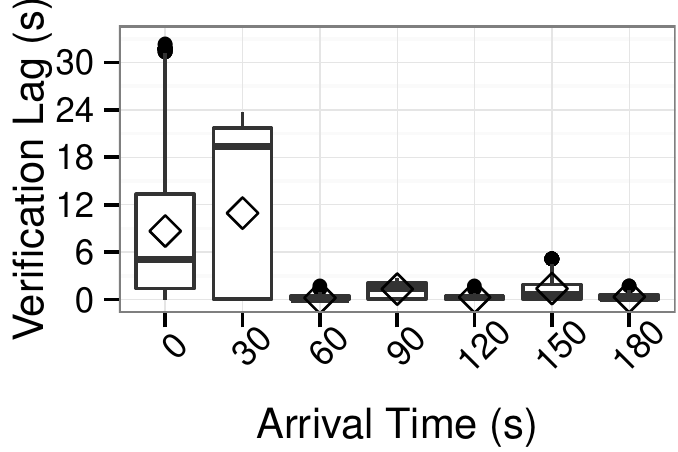}
}
\subfloat[][High-volume session]{
\label{fig:lag:elephant}
\includegraphics[width=0.5\columnwidth]{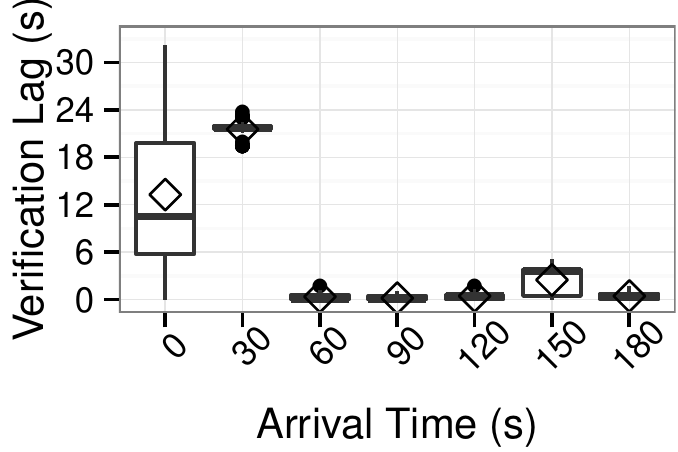}
}
\caption{Verification lags in seconds for the messages in the \gmail
  data set.  Box plot at horizontal-axis value \timestamp includes
  $\{\lag{\msgIdx}: \timestamp \le \arrival{\msgIdx} < \timestamp +
  30\secs\}$.  \figref{fig:lag:all} shows the verification lags for
  all messages in all 21 TLS sessions.  \figref{fig:lag:elephant}
  shows the verification lags for only the messages in the high-volume
  session.}
\label{fig:lag}
\end{figure}

%\begin{figure}[htb]
\begin{wrapfigure}{r}{0.55\columnwidth}
\centering
\includegraphics[width=0.5\columnwidth]{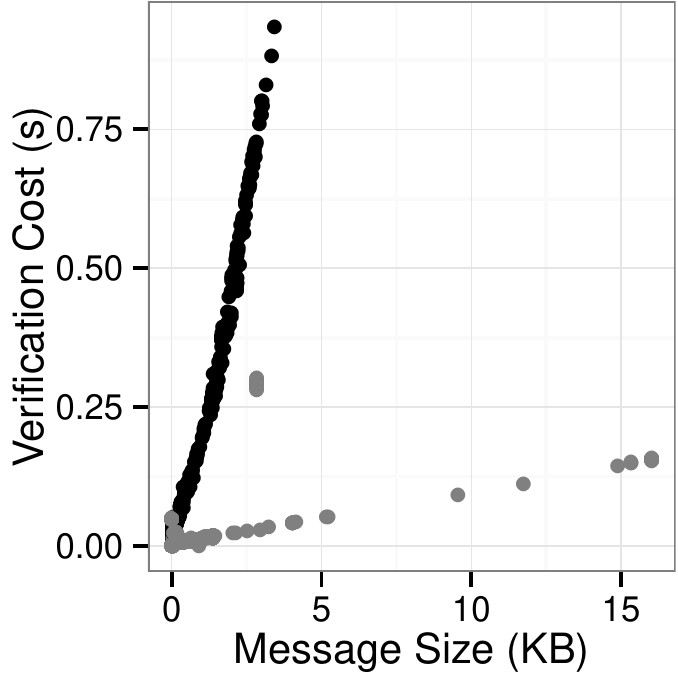}
\caption{Message size (kilobytes) versus verification cost (seconds).
  Client-to-server messages ($\bullet$) and server-to-client handshake messages
  (\textcolor{gray}{$\bullet$}) are shown.}
\label{fig:messagessize}
\end{wrapfigure}
%\end{figure}

\figref{fig:lag} shows the distribution of verification lag of
messages, grouped by the 30-second interval in which they arrived at
the verifier.
\figref{fig:lag:all} show all of the messages' verification lag,
whereas \figref{fig:lag:elephant} isolates the lag for the one
high-volume flow shown in \figref{fig:bw}.
It is evident from these figures that majority of the verification lag
happens early on, initially up to $\sim 30\secs$ in the worst case.
This lag coincides with an initial burst of
traffic related to requests while loading the \gmail application.
Another burst occurs later, roughly 160\secs into the trace, when an
attachment was uploaded by the client.  Still, even for the
high-volume session (\figref{fig:lag:elephant}), the lag is reduced to
near-zero in the middle of the session and by the end of the session.
The same holds true for the other 20 sessions, meaning that
verification for all sessions (in parallel) can easily complete within
approximately the wall-clock interval for which the sessions are
active.

\figref{fig:messagessize} shows verification cost as a function of
message size for all 21 TLS sessions.  Note that client-to-server
messages tend to be costlier to verify, despite being smaller, because
the verifier's execution of the client software when processing
server-to-client messages is almost entirely concrete.  In contrast,
the execution of the client in preparation of sending a
client-to-server message tends to involve more symbolic branching.
But what is most noteworthy about this plot is the linearity of the
relationship between message size and verification cost, particularly
for client-to-server messages. This is remarkable, given that the
general problem of client verification is undecidable.  In fact, it
provides a simple, application-independent way to estimate the costs
of verification for TLS sessions carrying payloads other than \gmail.
Furthermore, assuming similar upper bounds on the sizes of messages in
each direction, this predictability enables us to set a sharp timeout
at which point the verifier declares the client ``invalid.''  Based on
\figref{fig:messagessize}, setting the timeout at a mere 2\secs
(verification cost) would enable the verifier to quickly detect
misbehaving clients at a vanishingly small false alarm rate.

\begin{figure}[ht]
  %\centering
\hspace{-0.75em}
\subfloat[][All 21 TLS sessions]{
\label{fig:lag_drops2c:all}
\includegraphics[width=0.5\columnwidth]{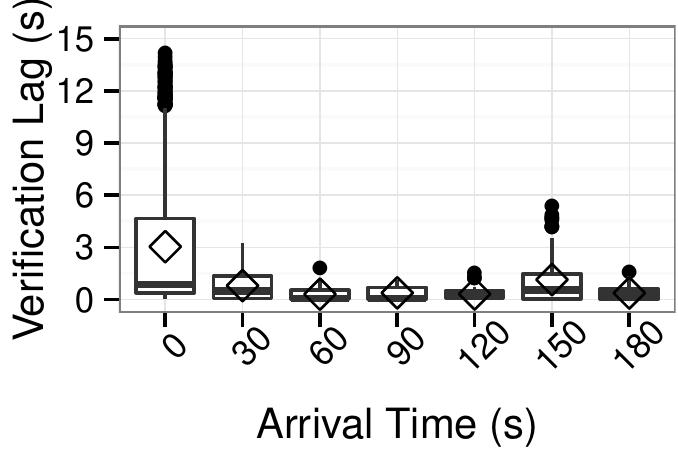}
}
\subfloat[][High-volume session]{
\label{fig:lag_drops2c:elephant}
\includegraphics[width=0.5\columnwidth]{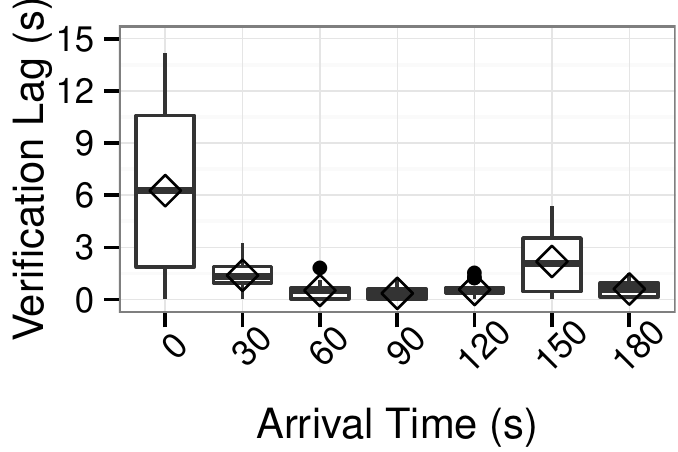}
}
\caption{Verification lags in seconds for the messages in the \gmail
  data set, dropping server-to-client application data messages.  Box
  plot at horizontal-axis value \timestamp includes $\{\lag{\msgIdx}:
  \timestamp \le \arrival{\msgIdx} < \timestamp + 30\secs\}$.
  \figref{fig:lag:all} shows the verification lags for all messages in
  all 21 TLS sessions.  \figref{fig:lag:elephant} shows the
  verification lags for only the messages in the high-volume session.}
\label{fig:lag_drops2c}
\end{figure}

\mypara{TLS-Specific Optimizations} While our goal so far in this
paper has been to provide for client behavior verification with a
minimum of protocol-specific tuning, a practical deployment should
leverage properties of the protocol for performance.  One important
property of TLS (and other TCP-based protocols such as SSH) is that
its client-to-server and server-to-client message streams operate
independently.  That is, with the exception of the initial session
handshake and ending session teardown, the verifiability of
client-to-server messages should be unaffected by which, if any,
server-to-client messages the client has received.  This gives the
verifier the freedom to simply ignore server-to-client application
data messages.  By doing so, the verification costs for
server-to-client messages, which effectively reduce to zero, do not
contribute to a growing lag.  The effect of this optimization on lag
is shown in \figref{fig:lag_drops2c}, in particular reducing the
worst-case lag to around $14\secs$. In all subsequent results, we have
ignored server-to-client messages unless otherwise noted.

\begin{figure}[ht]
%\centering
\hspace{-0.75em}
\subfloat[][$\workerCount = 1$]{
\label{fig:pad:t1pad128}
\includegraphics[width=0.5\columnwidth]{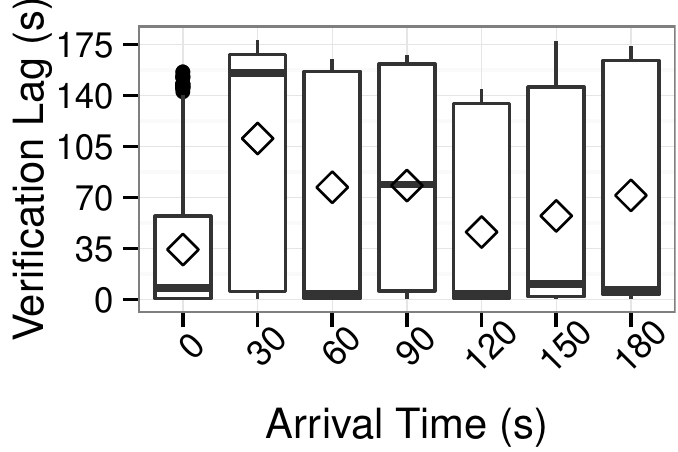}
}
\subfloat[][$\workerCount = 16$]{
\label{fig:pad:t16pad128}
\includegraphics[width=0.5\columnwidth]{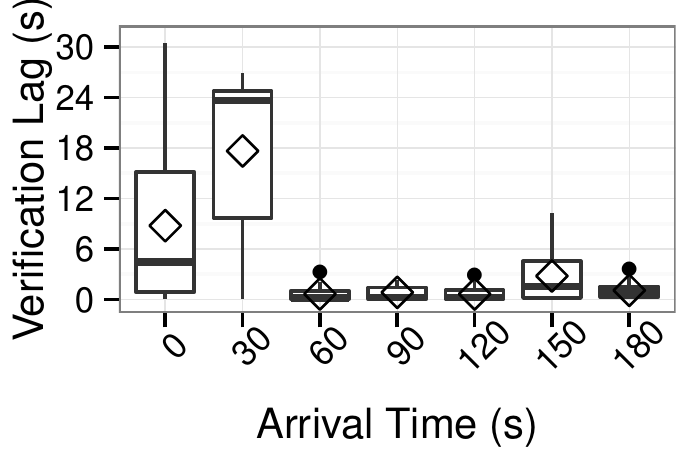}
}
\caption{Verification lags in seconds for the messages in the \gmail
  data set when up to 128 bytes of encrypted random padding are applied, over
  all 21 TLS sessions.  A box plot at horizontal-axis value \timestamp includes
  $\{\lag{\msgIdx}: \timestamp \le \arrival{\msgIdx} < \timestamp +
  30\secs\}$.  \figref{fig:pad:t1pad128}
  shows the lag for a one-worker verifier when padding is applied.
  \figref{fig:pad:t16pad128} shows the lag for a 16-worker verifier when
  padding is applied.}
\label{fig:pad}
\end{figure}

\subsection{Stress testing: Added protocol complexity}
\label{sec:eval:padding}

The \gmail performance evaluation showed that verification of a
typical TLS 1.2 session can be done efficiently and reliably, an
advance made possible by applying a multipass methodology to
cryptographic functions. In essence, once the state explosion from
cryptographic functions is mitigated, the client state space becomes
small enough that the verification time is primarily determined the
straight-line execution speed of the \klee symbolic interpreter.
However, not all clients are guaranteed to be this simple. One good
example is the draft TLS 1.3 standard~\cite{tls13}.  In order to hide
the length of the plaintext from an observer, implementations of TLS
1.3 are permitted (but not required) to pad an encrypted TLS record by
an arbitrary size, up to maximum TLS record size.  This random
encrypted padding hides the size of the plaintext from any observer,
whether an attacker or a verifier. In other words, given a TLS 1.3
record, the length of the input (e.g., from \texttt{stdin}) that was
used to generate the TLS record could range anywhere from 0 to the
record length minus header.  Other less extreme examples of padding
include CBC mode ciphers, and the SSH protocol, in which a small
amount of padding protects the length of the password as well as
channel traffic.

We thus extended our evaluation to stress test our verifier beyond
typical current practice.  We simulated the TLS 1.3 padding feature
by modifying a TLS 1.2 client (henceforth designated as ``TLS 1.2+''),
so that each TLS record includes a random amount of padding up to 128
bytes\footnote{While 128 bytes of padding may seem extreme, previous
  work showed that an attacker could sometimes infer the website
  visited by encrypted HTTP connections even with substantial padding
  (e.g.,~\cite{liberatore06:inferring}).}, applied before encryption.
We then measured the performance of our verifier, ignoring
server-to-client messages (except during session setup and teardown)
as before.

%\begin{figure}[htb]
\begin{wrapfigure}{r}{0.55\columnwidth}
  \begin{minipage}{0.55\columnwidth}
    \includegraphics[width=\textwidth]{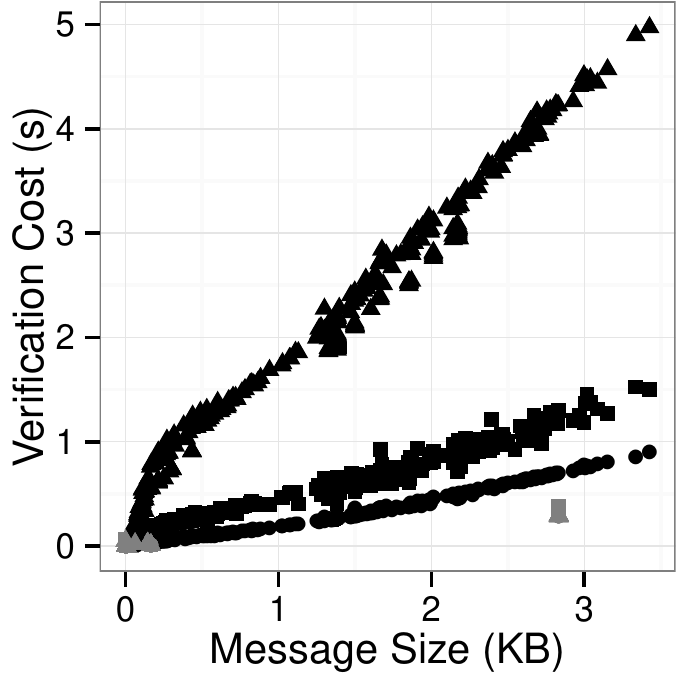}
    \vspace{-2ex}
  \caption{Message size (kilobytes) versus verification cost
    (seconds).  Shown are the single-worker baseline verifier running
    on TLS 1.2 traffic ($\bullet$), the single-worker verifier running
    on TLS 1.2+ ($\blacktriangle$), and the 16-worker verifier running
    on TLS 1.2+ ($\blacksquare$). TLS 1.2+ includes encrypted random
    padding. Client-to-server messages (black) and server-to-client
    handshake messages (gray) are also designated.}
  \label{fig:pad:messagessize}
  \end{minipage}
  \begin{minipage}{0.55\columnwidth}
  \vspace{3ex}
  \includegraphics[width=\textwidth]{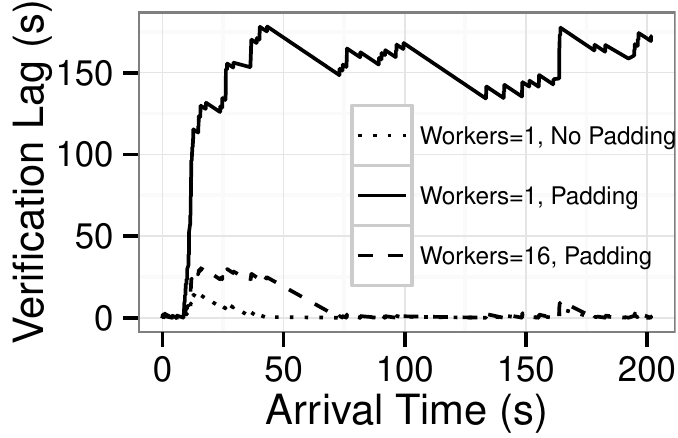}
  \vspace{-2ex}
  \caption{Verification lag in seconds for the messages in the
  high-volume session of the \gmail data set with $\leq 128$ bytes of
  encrypted random padding.}
  \label{fig:pad:trace1lags}
  \end{minipage}
\end{wrapfigure}
%\end{figure}

\figsref{fig:pad}{fig:pad:messagessize} show the
performance of our single- and 16-worker verifiers on TLS 1.2+
with a random amount of encrypted
padding.  \figref{fig:pad:messagessize} shows the verification cost as
a function of message size over all 21 \gmail TLS sessions and, for
comparison, the cost of our baseline single-worker verifier running on
a TLS 1.2 client as previously discussed in
\secref{sec:evaluation:gmail}.  The addition of random padding to TLS 1.2+
significantly enlarges the client state space that must
be explored.  When the single-worker verifier is applied to TLS 1.2+,
the verification cost increases substantially compared to
the TLS 1.2 baseline.  The 16-worker verifier reduces the verification
cost nearly back to the TLS 1.2 baseline levels.  This demonstrates
that the state space search is highly amenable to parallelization.
Again, it is noteworthy, and perhaps surprising, that in all three
cases, the verification cost is linearly related to message size.

\figref{fig:pad:trace1lags} shows the verification lag on a TLS 1.2+
client that is sending the high-volume \gmail trace (worst
case). Compared to the single-worker baseline on TLS 1.2, the
single-worker verifier on TLS 1.2+ is unable to keep up with the \gmail
high-volume traffic load, as its verification lag continues to
increase through the end of the session at which point it is over 4
minutes behind.  But with 16 workers, the initial lag (maximum of
$33\secs$ lag) reduces to near-zero by the $75\secs$ mark and the
verifier stays in step with the network traffic for the rest of the
session.  The 16-worker verifier parallelizes the search well enough
to keep up with a client whose state space is two orders of magnitude
larger than that of the typical TLS 1.2 client.

\section{Discussion}
\label{sec:discussion}

Here we discuss the applications for which our design is appropriate,
and several limitations of our approach.

\subsection{Applications}
\label{sec:discussion:applications}

\mypara{Suitable Application Layers} The application layer chosen for
our TLS evaluation, \gmail, was a relatively strenuous one in that it
exhibited both low-latency interactive behavior as well as high-volume
data transfer, the latter of which was challenging as measured by
verification lag.  Other applications that exhibit only one behavior
or the other may actually be more amenable to symbolic client
verification, not only as an intrusion detection system, but as an
intrusion \textit{prevention} system.\footnote{Because of the need for
  the verifier to obtain the master secret from the server after the
  server receives the second client-to-server message in the TLS
  handshake, the verifier would need to substitute a fake master
  secret in place of the real one in order to verify the first two
  client-to-server messages, if used as an intrusion prevention
  system.  The first \textit{use} of the real master secret, which is
  not evidenced until several messages later, would then cause the
  verifier to backtrack, at which point the real master secret could
  be inserted.}  For example, the Extensible Messaging and Presence
Protocol (XMPP) \cite{rfc6120}, generally used for text-based Internet
messaging, is highly interactive but sends relatively small XML
payloads.  An average verification cost of 126\msecs per TLS record may
be an acceptable latency for XMPP.  On the other end of the spectrum
are bulk data transfer applications such as the Simple Mail Transfer
Protocol (SMTP).  In this case, although the volume of data to be
transferred can be large, the application is highly tolerant of delay.
RFC 5321 recommends that the retry interval be at least 30 minutes and
the give-up time be at least 4-5 days \cite{rfc5321}.  Therefore, a
larger verification lag may be perfectly acceptable for TLS-protected
SMTP connections.

\mypara{Other Cryptographic Protocols}  Perhaps due to its widespread
use in various applications, TLS is one of the more complex security
protocols.  We believe that the client verification technique would
likely generalize to other, often simpler, protocols.  One example
is Secure Shell (SSH), which includes both authentication and
transport layer protocols~\cite{rfc4252,rfc4253}.  Indeed, SSH is
likely to exhibit application behavior amenable to our verification
approach.  When used as a remote shell, SSH requires low latency but
transfers a relatively small amount of data: key presses and terminal
updates.  When used for file transfer (SFTP), a large volume of data
is sent, but in a mode that is relatively latency-insensitive.

\subsection{Limitations}
\label{sec:discussion:limitations}

\mypara{Source Code and Other TLS Implementations}  Our verifier
requires the client source code to generate LLVM bitcode and to
designate prohibitive functions.  We also require knowledge of the
client configuration, such as command line parameters controlling the
menu of possible cipher suites.  While this work provides verification
for \openssl-based clients, there are several other popular
implementations of TLS, including Mozilla NSS, GnuTLS, Apple Secure
Transport, and Microsoft Schannel.  In principle, one could verify a
significant portion of TLS traffic by verifying observed traffic
against these five major TLS libraries, considering a client valid if
its behavior is consistent with any one of them.

\mypara{Client Versions} A verifier that is validating a client's
behavior against the wrong version of the client's implementation
could reject a legitimate client falsely.  In
\appref{sec:multi-version}, we summarize our initial steps toward
verifying multiple client versions far more efficiently than
instantiating a separate verifier for each possible version.

\mypara{Environment Modeling} While \openssl \sclient has relatively
few interactions with the environment, other clients may interact with
the environment extensively.  For example, \xpilot interacts with the
windowing system, and SSH reads \texttt{/etc/passwd}, the
\texttt{.ssh/} directory, redirects standard file descriptors,
etc. The \klee~\cite{cadar08:klee} and
\cloudnine~\cite{bucur11:cloud9} POSIX runtimes serve as good starting
points, but some environment modeling is likely to be necessary for
each new type of client. This one-time procedure per client is
probably unavoidable.

\mypara{Manual Choice of Prohibitive Functions}  We currently choose
the set of prohibitive functions manually.  While the choice of
cryptographic hash functions, public key algorithms, and symmetric
cipher primitives may be relatively obvious to a security researcher,
this may not be true for a typical software developer.

\mypara{Prohibitive Function Assumptions} When prohibitive functions
are initially skipped but eventually executed concretely, verification
soundness is preserved. If a prohibitive function is never executed
concretely (e.g., due to asymmetric cryptography), this introduces an
assumption; e.g., in the case of ECDH, a violation of this assumption
could yield an invalid curve attack~\cite{jager:2015:invalid}. In a
practical deployment, the user designating a prohibitive function
should also designate predicates on the function's output (e.g., the
public key is actually a group element) that are specified by the
relevant NIST or IETF standards as mandatory server-side
checks~\cite{rfc6090} (which would have prevented the Jager et
al.\ attack~\cite{jager:2015:invalid}). In our tool, these predicates
could be implemented via lazy constraint generation (see
\appref{sec:constraint-gen}), or as a \texttt{klee\_assume} for
simple predicates. Of course, care must be taken not to turn our
verifier into an oracle that enables Bleichenbacher-type
attacks~\cite{bleichenbacher:1998:cca}.

\mypara{IDS/IPS Deployment and Denial of Service}  Our current tool
takes as input a recorded network log.  To deploy it as an intrusion
detection system (IDS), it would need to be integrated with a passive
network tap.  For latency-tolerant applications, it could even be
deployed as an intrusion prevention system (IPS), where it acts as a
firewall that delivers only verified packets to the server.  In both
modes, but especially the latter, we must mitigate a potential denial
of service (DoS) attack. To do so, we leverage the strong linear
relationship between verification cost and message size in two ways.
(1) Impose a hard upper bound on verifier time per packet, and declare
all packets that exceed the time budget invalid. Since legitimate
packets finish within a few seconds, the bound can be easily set such
that the false alarm rate is negligible. (2) Given a fixed CPU time
budget, precisely compute the amount of traffic that can be
verified. The operator can then allocate verifiers according to the
threat profile, e.g., assigning verifiers to high-priority TLS
sessions or ones from networks with poor reputation
(e.g.,~\cite{collins07:uncleanliness}).  The IDS/IPS will then degrade
gracefully when total traffic bandwidth exceeds the verification
budget.

\section{Conclusion}
\label{sec:conclusion}

We showed that it is possible to efficiently verify that the messaging
behavior of an untrusted cryptographic client is consistent with
legitimate client code. Our technical contributions are twofold.
First, we described a parallel client verification engine that
supports concurrent exploration of paths in the client software to
explain a sequence of observed messages.  This innovation is both
generally useful for client verification and specifically useful for
verifying cryptographic clients, e.g., due to ambiguities arising from
message padding hidden by encryption.  Second, we developed a
multipass verification strategy that enables verification of clients
whose code contains cryptographic functions, which typically pose
major challenges to symbolic execution.  We demonstrated our verifier
by showing that it detects two attacks on \openssl that represent two
classes of client misbehavior: those that produce malformed messages
(e.g., Heartbleed), and those that leverage correctly formatted
messages that are nevertheless impossible to observe from a valid
client.  In addition, we showed that our verifier can keep pace with a
typical TLS load (\gmail), running over both \openssl TLS 1.2 and a
more complex simulation of TLS 1.3.  We believe our contributions to
be significant in reducing the detection time of a nontrivial class of
protocol exploits from years to seconds, with no prior knowledge of
the exploit.

\mypara{Acknowledgments}
This work was supported in part by NSF grants 115948 and 1330599,
grant N00014-13-1-0048 from the Office of Naval Research, and a gift
from CISCO.

{\footnotesize \bibliographystyle{acm}
\bibliography{paper}}

\appendix

\section{Demonstration of parallelization on client-server games}
\label{sec:parallel-demo}

Though our focus in this paper is on cryptographic protocols, the
benefits of the parallel verification architecture described
in \secref{sec:parallel} extend to other protocols, as well.  To
demonstrate this, we employ two game clients studied in previous
work~\cite{cochran13:verification}, namely
\xpilot and \tetrinet (specifically \xpilotng v4.7.2 and \tetrinet
v0.11).  The \xpilot client consists of roughly 100,000 SLOC.  Beyond
this, the scope of symbolic execution included all needed libraries
except \xlib, whose functions were replaced with minimal stubs, so
that the game could be run without display output.  Moreover, \uclibc
was used in lieu of the GNU C library.  The \tetrinet client is 5000
SLOC.  As in \xpilot, the scope of symbolic execution also included
all needed libraries, though again the display output library
(\ncurses) was disabled using minimal stub functions and \uclibc was
used in place of the GNU C library.  The experiments shown here were
run on a system with $256\gbytes$ of RAM and $3.2\ghertz$ processor
cores.  Execution fragments were prioritized by \selectNode as
described in prior work~\cite{cochran13:verification}.

\begin{figure}[t]
%\centering
\hspace{-0.75em}
\subfloat[][$\workerCount = 1$]{
\label{fig:tetrinet:lag:parallel_1}
\includegraphics[width=0.5\columnwidth]{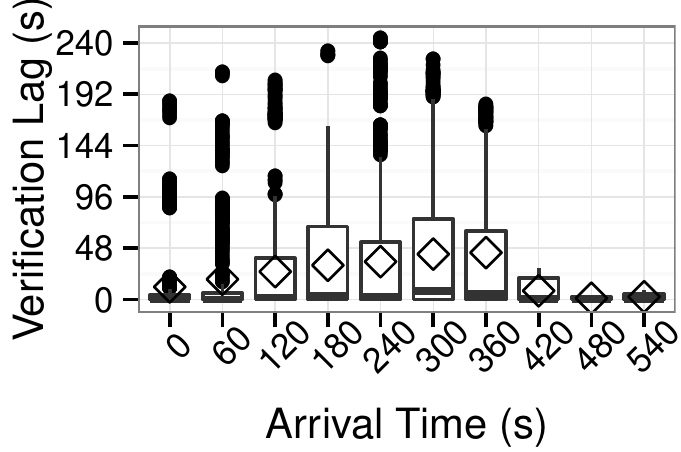}
}
\subfloat[][$\workerCount = 16$]{ 
\label{fig:tetrinet:lag:parallel_16}
\includegraphics[width=0.5\columnwidth]{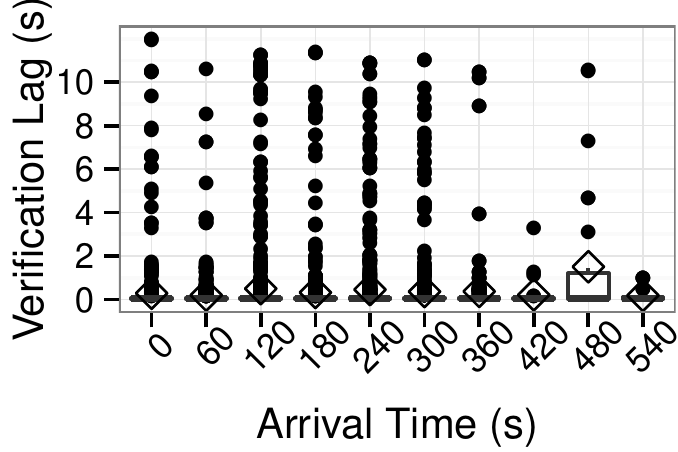}
}
\caption{Verification lag of legitimate \tetrinet traces.  Box plot at
  horizontal-axis value \timestamp includes $\{\lag{\msgIdx}:
  \timestamp \le \arrival{\msgIdx} < \timestamp + 60\secs\}$ in
  \tetrinetTraces traces, each starting at time $\timestamp=0$.
  ``$\Diamond$'' shows the average.}
\label{fig:tetrinet:parallel:lag}
\end{figure}

\figref{fig:tetrinet:parallel:lag} shows the distribution of
verification lag per message, binned into $60\secs$ bins, across
\tetrinetTraces{} \tetrinet traces.  The boxplot labeled \timestamp
shows the distribution of verification lags for messages that arrived
between times \timestamp and $\timestamp+60\secs$ in the
\tetrinetTraces traces.  In each boxplot, the ``box'' shows the first,
second (median) and third quartiles, and its whiskers extend to cover
points within $\pm 1.5$ times the interquartile range.  Additional
outlier points are shown as dots.  Overlaid on each boxplot is a
diamond ($\Diamond$) that shows the average of the data points.  In
the single worker configuration
(\figref{fig:tetrinet:lag:parallel_1}), verification lags behind
message arrival times by more than $200\secs$ in the worst case.  In
contrast, with 16 workers (\figref{fig:tetrinet:lag:parallel_16}),
verification is able to keep pace with gameplay and never accumulates
lag over the course of verification.  Even if verification falls
behind at some point in the game, it always catches up because of the
gap between message arrival times.  As such, the verifier should need
only a fixed sized buffer of network messages to manage a long-running
verification session.

\begin{figure}[t]
%\centering
\hspace{-0.75em}
\subfloat[][$\workerCount = 1$]{
\label{fig:xpilot:lag:parallel_1}
\includegraphics[width=0.5\columnwidth]{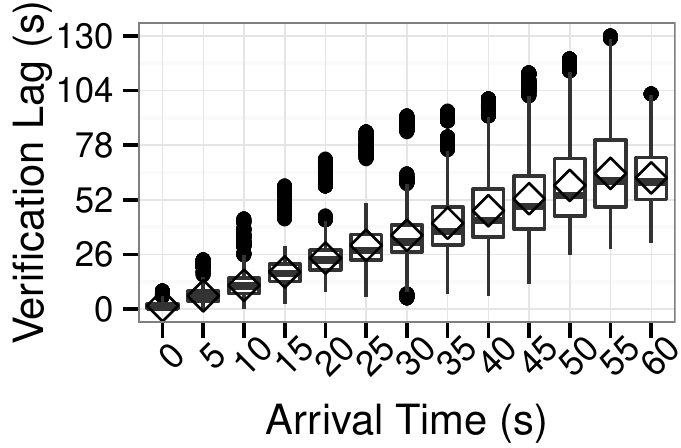}
}
\subfloat[][$\workerCount = 8$]{
\label{fig:xpilot:lag:parallel_8}
\includegraphics[width=0.5\columnwidth]{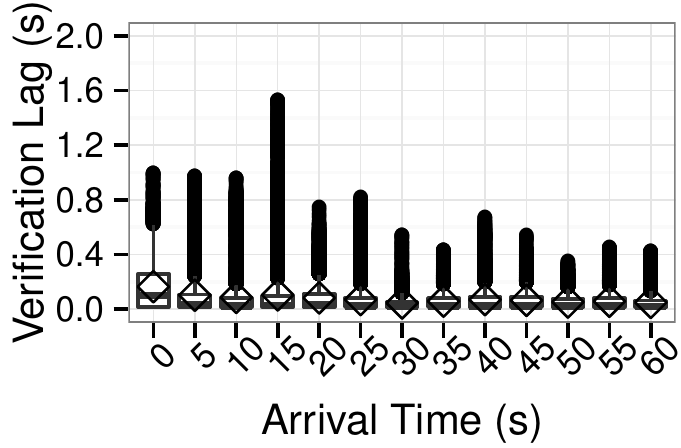}
}
\caption{Verification lag of legitimate \xpilot traces.  Box plot at
  horizontal-axis value \timestamp includes $\{\lag{\msgIdx}:
  \timestamp \le \arrival{\msgIdx} < \timestamp + 5\secs\}$ in
  \xpilotTraces traces, each starting at time $\timestamp=0$.
  ``$\Diamond$'' shows the average.}
\label{fig:xpilot:parallel:lag}
\end{figure}

In \figref{fig:xpilot:parallel:lag}, the verification lags for \xpilot
are shown for two worker configurations across \xpilotTraces
message traces.  Despite a mean verification cost of $75\msecs$ when
using a single worker thread, the fast pace of \xpilot makes it
difficult for verification to keep pace with the game
(\figref{fig:xpilot:lag:parallel_1}).  However, by increasing the
number of worker threads, we can see that in the 8-worker
configuration (\figref{fig:xpilot:lag:parallel_8}), verification lag
never significantly falls behind and could use only a fixed buffer of
messages for verification in long-running sessions.  In this case,
there is little additional improvement gained by moving to 16 workers.

\section{TLS Experimental Setup}
\label{sec:tls-setup}

In \secref{sec:evaluation}, we applied our client verification
algorithm to \openssl, a widely used implementation of Transport Layer
Security (TLS) with over 400,000 lines of code. We configured our
\openssl client with one of the currently preferred cipher suites, namely
{\small \texttt{TLS\_ECDHE\_ECDSA\_WITH\_AES\_128\_GCM\_SHA256}}.
\begin{itemize}
\item Key exchange: Ephemeral Elliptic-Curve Diffie Hellman (ECDHE)
  signed using the Elliptic Curve Digital Signature Algorithm (ECDSA)
\item Symmetric Encryption: 128-bit Advanced Encryption Standard (AES) in
  Galois/Counter Mode
\item Pseudorandom function (PRF) built on SHA-256
\end{itemize}

Since our goal was to verify the TLS layer and not the higher-layer
application, in our experiment we took advantage of the \openssl
\sclient test endpoint.  This client establishes a fully
functional TLS session, but allows arbitrary application-layer data to
be sent and received via \texttt{stdin} and \texttt{stdout}, similar
to the \texttt{netcat} tool. Verifying that network traffic is
consistent with \sclient is roughly equivalent to verifying the
TLS layer alone.

The \openssl-specific user configuration for verification consisted of
the following:
\begin{enumerate}
\item Configuring the following \openssl functions as prohibitive:
  {\small \tt AES\_encrypt, ECDH\_compute\_key,\\ EC\_POINT\_point2oct,
  EC\_KEY\_generate\_key,\\ SHA1\_Update, SHA1\_Final, SHA256\_Update,\\
  SHA256\_Final, gcm\_gmult\_4bit, gcm\_ghash\_4bit}
\item Configuring \texttt{\small tls1\_generate\_master\_secret} as the function
  to be replaced by server-side computation of the symmetric key.
\item (Optional) Declaring \texttt{\small EVP\_PKEY\_verify} to be a function
  that always returns success. Performance optimization only.
\end{enumerate}
The user configuration for \openssl, comprising declarations of
prohibitive functions and their respective input/output annotations,
consisted of 138 lines of C code using our
API.  \figref{fig:prohib:aes} shows an example prohibitive function
declaration for the AES block cipher.  In this macro, we declare the
function signature, which comprises the 128-bit input
buffer \texttt{in}, the 128-bit output buffer \texttt{out}, and the
symmetric key data structure, \texttt{key}, which contains the
expanded round keys. Both \texttt{in} and \texttt{key} are checked for
symbolic data.  If either buffer contains symbolic data, \texttt{out}
is populated with unconstrained symbolic data, and the macro returns
without executing any subsequent lines.  Otherwise, the underlying (concrete)
AES block cipher is called.

\newbox\aesbox
\begin{lrbox}{\aesbox}
\lstset{numbers=none,basicstyle=\small\ttfamily}
\begin{lstlisting}
DEFINE_MODEL(void, AES_encrypt,
             const unsigned char *in,
             unsigned char *out,
             const AES_KEY *key)
{
  SYMBOLIC_CHECK_AND_RETURN(
      in, 16,
      out, 16, "AESBlock");
  SYMBOLIC_CHECK_AND_RETURN(
      key, sizeof(AES_KEY),
      out, 16, "AESBlock");
  CALL_UNDERLYING(AES_encrypt,
      in, out, key);
}
\end{lstlisting}
\end{lrbox}

\begin{figure}[t]
\usebox\aesbox
\caption{Example prohibitive function declaration.
\label{fig:prohib:aes}}
\end{figure}

In a pure functional language or an ideal, strongly typed language,
the prohibitive function specifications could in principle be
generated automatically from the function name alone.  Unfortunately,
in C, the memory regions representing input and output may be
accessible only through pointer dereferences and type casts.  This is
certainly true of \openssl (e.g., there is no guarantee that
the \texttt{AES\_KEY} struct does not contain pointers to auxiliary
structs).  Therefore, for each prohibitive function, the user
annotation must explicitly define the data layout of the input and
output.

The domain knowledge required for the first two configuration steps is
minimal, namely that current TLS configurations use the above
cryptographic primitives in some way, and that a symmetric key is
generated in a particular function.  The domain knowledge necessary
for the third configuration step is that TLS typically uses public key
signatures only to authenticate the server to the client, e.g., via
the Web PKI.  The server itself generates the signature that can be
verified via PKI, and so the verifier knows that the chain of
signature verifications is guaranteed to succeed.  Moreover, this
optimization generalizes to any protocol that uses a PKI to
authenticate the server to an untrusted client.

\section{Multi-version Verifiers}
\label{sec:multi-version}

When the version of the client software used by the verifier is
different from the version of the client software being run by a
legitimate client, it is possible for the verifier to falsely accuse
the client as being invalid.  This poses a challenge for verification
when the client version is not immediately evident to the verifier.
For example, \openssl does not communicate the minor version number of
its client to the server, and hence our verifier would be in the dark
about this version number.  The possibility for false accusations here
is real: we confirmed, for example, that a verifier for \openssl
client version \sslvE can fail if used to verify traffic for \openssl
client version \sslvF, and vice versa.  This occurs because, for
example, the changes from \sslvE to \sslvF included removing MD5 from
use and removing a timestamp from a client nonce, not to mention
numerous platform-specific adaptations and bug fixes.  In total,
\sslvF involved changes to 102 files amounting to 1564 insertions and
997 deletions (according to \git), implemented between Feb 11, 2013
and Jan 6, 2014.

The immediate solution to this problem is to instantiate a verifier
for every possible version that a legitimate client might be using.
By running these verifiers in parallel on the message trace from the
client, the trace can be considered valid as long as one verifier
remains accepting of the trace.  Running many verifiers in parallel
incurs considerable expense, however.

An alternative approach to address this issue is to create a single
verifier that verifies traffic against several versions simultaneously
--- a \textit{multi-version verifier} --- while amortizing
verification costs for their common code paths across all versions.
To illustrate the potential savings, we constructed a multi-version
verifier for both \sslvE and \sslvF by manually assembling a ``unioned
client'' of these versions, say client ``\sslvEF''.  In client
\sslvEF, every difference in the code between client \sslvE and client
\sslvF is preceded by a branch on version number, i.e.,
\begin{verbatim}
if (strcmp(version, "1.0.1e") == 0) { 
   /* 1.0.1e code here */
} else { 
   /* 1.0.1f code here */
}
\end{verbatim}
We then provided this as the client code to the verifier, marking
\verb|version| as a symbolic variable.  It is important to note that
once the client messages reveal behavior that is consistent with only
one of \sslvE and \sslvF, then \verb|version| will become concrete,
causing the verifier to explore only the code paths for that version;
as such, the verifier still allows only ``pure \sslvE'' or ``pure
\sslvF'' behavior, not a combination thereof.

As shown in \tblref{multi-version:overhead}, the single-worker costs
(specifically, $\sum_{\msgIdx} \cost{\msgIdx}$) of verifying \sslvE
traffic with a \sslvEF verifier and of verifying \sslvF traffic with a
\sslvEF verifier were both within 4\% of the costs for verifying with a
\sslvE and \sslvF verifier, respectively.  (For these tests, we used
the same \gmail traces used in \secref{sec:evaluation}.)  In fact, we
do not include lag graphs for the multi-version verifier here
(analogous to those in \secref{sec:evaluation}) because they are
visibly indistinguishable from those for single-version verifiers. As
can be seen from \tblref{multi-version:overhead}, despite a 32\%
increase in symbolic branches that in turn drives a 7\% increase in
SMT solver queries, the overall cost increases very little. This
implies that despite an increase in the number of code path
``options'' comprising the union of two versions of client code, the
incorrect paths die off quickly and contribute relatively little to
total verification cost.

While a demonstration of the efficiency of a multi-version verifier
for only two versions of one codebase, we believe this result suggests
a path forward for verifying clients of unknown versions much more
efficiently than simply running a separate verifier for each
possibility.  We also anticipate that multi-version verifiers can be
constructed automatically from commit logs to repositories, a
possibility that we hope to explore in future work.

\begin{table}[tbp]
\centering
\begin{tabular}{ |c||c|c|  }
 \hline
 &\multicolumn{2}{|c|}{Network Trace} \\
 \hline
 Measurement & \sslvE & \sslvF \\
 \hline
 Symbolic branches    &        32.3\%  &        32.0\% \\
 SMT solver queries   &         7.8\%  &         7.6\%  \\
 \textbf{Verification cost} & \textbf{3.0\%} & \textbf{3.3\%} \\
 \hline
\end{tabular}
\caption{Percentage overhead incurred by verification with a ``union''
       \sslvEF client instead of the matching
       client. Verification cost (time) is defined
       in \secref{sec:evaluation}.  \label{multi-version:overhead}}
\end{table}

\section{Extension: Lazy Constraint Generators}
\label{sec:constraint-gen}

There are several potentially useful extensions to our
client verification algorithm that we are considering for future
development.  Here we highlight one, namely \textit{lazy constraint
  generators} to accompany the designation of prohibitive functions.
Since a function, once specified as prohibitive, will be skipped by
the verifier until its inputs are inferred concretely, the verifier
cannot gather constraints relating the input and output buffers of
that function until the inputs can be inferred via other constraints.
There are cases, however, where introducing constraints relating the
input and output buffers once some subset of them are inferred
concretely would be useful or, indeed, is central to eventually
inferring other inputs concretely.

Perhaps the most straightforward example arises in symmetric
encryption modes that require the inversion of a block cipher in order
to decrypt a ciphertext (e.g., CBC mode).  Upon reaching the client
\sendInstr instruction for a message, the verifier reconciles the
observed client-to-server message \msg{\msgNmbr} with the constraints
\newState.\constraints accumulated on the path to that \sendInstr;
for example, suppose this makes concrete the buffers corresponding to
outputs of the encryption routine.  However, because the block cipher
was prohibitive and so skipped, constraints relating the input buffers
to those output buffers were not recorded, and so the input buffers
remain unconstrained by the (now concrete) output buffers.  Moreover,
a second pass of the client execution will not add additional
constraints on those input buffers, meaning they will remain
unconstrained after another pass.

An extension to address this situation is to permit the user to
specify a lazy constraint generator along with designating the block
cipher as prohibitive.  The lazy constraint generator would simply be
a function from some subset of the input and output buffers for the
function to constraints on other buffers.  The generator is ``lazy''
in that it would be invoked by the verifier only after its inputs were
inferred concretely by other means; once invoked, it would produce new
constraints as a function of those values.  In the case of the block
cipher, the most natural constraint generator would be the inverse
function, which takes in the key and a ciphertext and produces the
corresponding plaintext to constrain the value of the input buffer.

Our \openssl case study in \secref{sec:multipass:tls} does not require
this functionality since in the encryption mode used there, the
ciphertext and plaintext buffers are related by simple exclusive-or
against outputs from the (still prohibitive) block cipher applied to
values that can be inferred concretely from the message.  So, once the
inputs to the block cipher are inferred by the verifier, the block
cipher outputs can be produced concretely, and the plaintext then
inferred from the concrete ciphertexts by exclusive-or.

\end{document}